\documentclass[AMA,STIX1COL]{WileyNJD-v2}
\articletype{Article Type}%
\usepackage{amsmath}
\received{26 April 2016}
\revised{6 June 2016}
\accepted{6 June 2016}
\usepackage{makecell}
\usepackage{subcaption}
\usepackage[font=large]{caption}
\usepackage{graphicx}
\usepackage[most]{tcolorbox}

\usepackage{xcolor}

\begin{document}

\title{Stochastic Geometry of Network \\of Randomly Distributed Moving Vehicles\\ on a Highway}

\author[]{Gleb Dubosarskii}

\author[]{Serguei Primak}

\author[]{Xianbin Wang}

\authormark{Gleb Dubosarskii \textsc{et al}}

\corres{Gleb Dubosarskii, Department of Electrical and Computer Engineering, Western University, Ontario, Canada. \email{gdubosar@uwo.ca}}


\abstract[Summary]{Vehicular ad-hoc networks (VANETs) have become an extensively studied topic in contemporary research.
One of the fundamental problems that has arisen in such research is understanding the network statistical properties, such as the cluster number distribution and the cluster size distribution.
In this paper, we analyze these characteristics in the case in which vehicles are located on a straight road.
Assuming the Rayleigh fading model and a probabilistic model of intervehicle distance,
we derive probabilistic distributions of the aforementioned connectivity characteristics, as well as distributions of the biggest cluster and the number of disconnected vehicles.
All of the results are confirmed by simulations carried out for the realistic values of parameters.}


\jnlcitation{\cname{%
\author{Dubosarskii G.},
\author{S. Primak}, and
\author{X. Wang}} (\cyear{2019}),
\ctitle{Stochastic Geometry of Network of Randomly Distributed Moving Vehicles on a Highway}, \cjournal{Intern. J. Commun. Systems}, \cvol{2019;00:1--6}.}

\maketitle


\section{Introduction}
Vehicular ad-hoc networks (VANETs) are developed to provide intelligent transportation that improves road safety, reduces transportation time, and decreases road congestion. One of the possible signal propagation scenarios is vehicle-to-vehicle (V2V) communication, which means a direct connection between two nearby vehicles. Another considered scenario is vehicle-to-infrastructure (V2I) communication, in which a vehicle sends and receives a signal from Road Side Units (RSUs) at the side of the road. RSUs  form infrastructure that improves the performance of VANETs and enhances information distribution.

Multihop information propagation in VANET is limited by high vehicular speed and frequent disconnections. To address these issues, significant progress was made in routing protocols \cite{R1,R2,R3,R4,R5} developed to reduce latency and stabilize the connection between vehicles, allowing vehicles to receive real-time travel-related information.

The high mobility of vehicles and the changing network topology lead to difficulties in the distribution of content, including images and video. To mitigate these shortcomings, cloud architectures are proposed that allow for the sharing of abundant resources, and for the regulating of network connectivity specifications. In \cite{Cl1}, the scheme, in which parked vehicles can play the role of RSUs, is proposed. In \cite{Cl2}, RSU cloud is introduced that has the ability to dynamically reconfigure data forwarding rules in the network to meet frequently changing service demands. Article \cite{Cl3} integrates both RSU and vehicles into one computational network that provides services, and proposes an optimal strategy for resource allocation.

VANET is a highly dynamic network, making it difficult to analyze from a deterministic point of view. This is the reason that statistical methods are used to make predictions under the assumption of particular probabilistic distributions of vehicles on the road and models of signal transmission. Considering this problem on a real map seems to be an impossible task, so, in most articles, the authors limit themselves to the cases of a highway or a crossroad. Articles \cite{1,2,3,Clust,ChModel} are devoted to the investigation of statistical characteristics of the network, such as the number of clusters, cluster size, and the number of disconnected vehicles. Let us discuss the content of these articles in more detail. In \cite{1}, it is assumed that the probabilistic distribution of distance between cars is known, and as soon as the distance between neighboring vehicles does not exceed transmission range $L$, they are always able to establish a connection.
Under these assumptions, probability distribution of information propagation distance is found, as well as its  expected value and variation. Articles \cite{2, ChModel} discuss more realistic communication channel models, such as Rayleigh, Rician, and Weibull fading channels. For each of these models, the probability of connection between neighboring vehicles and the probability of full connectivity of the network are derived. The authors of \cite{3} consider a more complex case in which the cars are located between two RSUs, but use a simple the model of the communication channel: they assume (as in \cite{1}) that the vehicles are always connected within the constant transmission range and disconnected otherwise. Such advanced network characteristics as expected number of clusters, network connectivity probability, expected
number of vehicles required for a VANET to be fully connected, and expected network capacity are found within the framework of this model. In \cite{Clust}, the authors suggest that vehicles in the cluster do not communicate with each other, but, instead, all vehicles communicate with the main car. Under presumption of the Rayleigh fading channel model, the Markov model of packet transmission, and the PHY decoding failure model, the average packet
loss probability is derived. In \cite{CarNet,CarNet2,Myev}, the evolution of the network over time is considered. In \cite{CarNet}, a case in which  a signal is transmitted to vehicles moving in the opposite direction and then sent back to the initial side is investigated. The probability of such a successful two-hop connection is derived under the assumption that vehicles move at a constant speed. Article \cite{CarNet2} focuses on the study of link duration in the specific case, in which the speed of the vehicles grows linearly, reaches the limit, and then remains constant. In our article \cite{Myev}, the network evolution is investigated under the presumption that the connection between consecutive cars is preserved according to the Markov probabilistic model; its parameters are explicitly expressed through the network macro parameters. Under these conditions, explicit formulas are obtained for the probability distribution of link duration between consequent vehicles, the cluster existence, and other fundamental network characteristics.


In this article, we consider the case of a vehicle network on a highway. Our goal is to give an extended description of the network, not only in terms of average values and variance, as is done in the previous articles (such as \cite{3}), but to find the whole distribution of the number of clusters, cluster size, and the number of disconnected vehicles (here we call them \textit{idle} vehicles). We use a more realistic model of the communication channel (Rayleigh fading channel) than in \cite{1,3}. We advance further than the authors of \cite{2,Clust, ChModel} and find the distribution of the number of clusters and cluster size. The difference between this article and \cite{CarNet, CarNet2, Myev} is that we do not consider the evolution of the network over time, but concentrate in detail on more nuanced connectivity characteristics at a fixed moment in time.
Moreover, to the best of our knowledge, we are the first to investigate the distribution of the biggest cluster size. We assume that every car communicates only with the nearest car, in front, and behind it. This assumption is made in the previous statistical research in this area, in order to make the model simple enough to analyze.  Under the presumption of  Rayleigh fading model and known density function of intervehicle distance, we express the above mentioned distributions in terms of known parameters of the network. Our model can be applied to other scenarios if the probability of connection between every pair of consecutive vehicles has the same value $p$. We derive several interesting results describing network connectivity, namely, that the average size of cluster is a constant, approximately equaling $1/(1-p)$; the average number of clusters is proportional to the number of vehicles $n$ equaling $1+(n-1)(1-p)$; and in the network, on average there is a constant fraction $\approx 1/(1-p)^2$ of the \textit{idle} vehicles. The average largest cluster of the network, however, is not a constant and grows as $\log_{1/p}n$. We confirm these and other theoretical results by simulations carried out in the cases of different car densities and number of cars. The averages and variances of the studied properties are summarized in the following table (where $\gamma=0.577\ldots$ is an Euler constant):
\begin{center}
\begin{tabular}{|c|c|c|}
\hline
Property    & Average & Variance \\
\hline
Number of clusters   & $1+(n-1)(1-p)$ & $(n-1)p(1-p)$ \\
\hline
Size of clusters   & $(1-p^n)/(1-p)$ & $(2p^{n+1}n-2p^n n-p^{2n}-p^{n+1}+p^n+p)/(1-p)^2$ \\
\hline
Size of the biggest cluster & $\approx \log_{1/p}\{(n-1)(1-p)\}+\gamma/\ln(1/p)+0.5$ & $\approx\pi^2/\ln^2(1/p)+\frac{1}{12}$ \\
\hline
Number of idle cars & $2(1-p)+(n-2)(1-p)^2$ &  \thead{$-3np^4 + 10np^3 - 11np^2 + 4np+$\\$ 8p^4 - 22p^3 + 18p^2 - 4p$}\\
\hline
\end{tabular}
\end{center}

The studied statistical characteristics have several practical applications. Cluster size distribution and maximum cluster size distribution provide us with tools for network load prediction. Also, this statistical information can be used for security purposes. The cluster size estimation, for example, allows for the prediction of the number of infected vehicles in the event of an attack on a cluster. Estimation of the number of disconnected vehicles makes it possible to evaluate the quality of connection, in order to make the percentage of disconnected vehicles acceptably low.

The paper is organized as follows. In section \ref{Sec:1} we describe the connectivity model and derive a formula for the probability of connection between two consecutive cars. In \ref{Sec:3} we derive all the probabilistic distributions mentioned above. Finally, in section \ref{Sec:4} the simulation results are presented and compared to the calculations done by the formulas from section \ref{Sec:3}. The appendix is devoted to a quick introduction to the theory of generating functions needed in section \ref{Sec:3}. We use the following abbreviations:


We use the following abbreviations for the connectivity characteristics of the network:
\begin{center}
\begin{tabular}{|c|c|}
\hline
ClustNum     & number of clusters in the network \\
ClustSize    & size of the cluster               \\
BiggestClust & size of the biggest cluster       \\
IdleCars     & number of disconnected vehicles   \\
\hline
\end{tabular}
\end{center}

The following variables and functions are used:
\begin{center}
\begin{tabular}{|c|c|}
\hline
  $p$ & probability of connection between two consecutive vehicles \\
  $G_T$ & transmit antenna gain\\
  $G_R$ & receive antenna gain\\
  $P_{tx}$ & transmit power \\
  $\alpha$ & path loss exponent\\
  $K$ & constant associated with the path loss model\\
  $d$ & distance between cars\\
  $C$ & speed of light\\
  $W$ & thermal noise power\\
  $k$ & Boltzmann constant, $k=1.38\times 10^{-23} J/K$\\
  $f_c$ & carrier frequency\\
  $T_0$ & room temperature\\
  $B$ & transmission bandwidth\\
  $f_\gamma(x)$ & Signal-to-noise ratio probability density function\\
  $f_d(x)$ & intervehicle distance probability density function\\
  $\binom{k}{s}$ & binomial coefficient \\
  $coeff_{x^n} f(x)$ & coefficient of the term $x^n$  in series $f(x)$\\
  $clust(N)$ & number of clusters of vehicle network $N$\\
  $count(r;N)$ & number of clusters of vehicle network $N$ having size $r$\\
  $Num(s,k)$ & number of networks  having $k$ clusters, under condition that $s$ of them have size $r$.\\
  $\gamma$ & Euler constant, $\gamma=0.577\ldots$\\
  \hline

\end{tabular}
\end{center}

\section{Network model} \label{Sec:1}

We consider the network of randomly distributed cars on a highway and suppose that every vehicle of the network can establish a connection only with the closest front and back neighbours.  We denote by $d$ the distance between two consecutive vehicles. We assume that the distance between vehicles has random distribution with density function $f_d(x)$. Typically it is assumed that distance between cars has
exponential distribution, normal distribution, gamma
distribution or log-normal distribution. These distributions correspond to different flow conditions. For instance, exponential distribution corresponds to low traffic flow conditions, while high traffic flow conditions are described by normal distribution.

Let us suppose that there are $n$ cars on a road and distribution of distance between cars has a mean value $M$, therefore, the average distance between the first and the last car is $(n-1)M$, consequently, despite the fact that the road is long, it is improbable that the distance between first and last vehicle is very large.

We assume that the average channel Signal-to-noise ratio (SNR) between two consecutive vehicles $\overline{\gamma}$ is calculated using the following formula from \cite{ChModel}:
\begin{equation}\label{11144}
\overline{\gamma}=\frac{P_{tx}K}{d^\alpha W},
\end{equation}
where $P_{tx}$ is the
transmit power, $\alpha$ is the path loss exponent and $K$ is a
constant associated with the path loss model. The parameter $K$ is given by the formula
\begin{equation}\label{123456}
K =\frac{G_T G_R C^2}{(4\pi f_c)^2},
\end{equation}
where $G_T$ and $G_R$ are the transmit and receive antenna gains, $C$ is the
speed of light and $f_c$ is the carrier frequency. We assume that the antennas are omni directional
\begin{equation}\label{1234567}
G_T = G_R = 1
\end{equation}
and $f_c=5.9 GHz$.
The thermal noise power can be determined by the formula
\begin{equation}\label{123456733}
W =k T_0 B,
\end{equation}
where $k =1.38\times 10^{-23} J/K$ is the Boltzmann constant, $T_0$ ($T_0 =300^\circ K$) is the room
temperature, and
$B$ ($B = 10$ MHz) is the transmission bandwidth.



The SNR density function $f_\gamma(x)$ (is different from $f_d$) is given by the formula
\begin{equation}\label{11144x}
f_\gamma(x)=\frac{1}{\overline{\gamma}}e^{-x/\overline{\gamma}}.
\end{equation}
It is assumed that there is a connection between consecutive cars if $\gamma$ is greater than the threshold $\Psi$. Therefore,
\begin{equation}\label{11144d}
P(\gamma>\Psi)=\int_\Psi^\infty f_\gamma(x)dx=e^{-\Psi/\overline{\gamma}}.
\end{equation}

Let us denote the probability that two consecutive vehicles can establish a connection by $p$.
By using (\ref{11144}), (\ref{11144d}) and following the logic of \cite{3} we can derive formula expressing value of the probability $p$ through intervehicle density function $f_d(x)$:
\begin{equation}\label{ppppp}
p=\int_0^\infty P(\gamma(x)>\Psi)f_d(x)dx= \int_0^\infty e^{-\frac{\Psi x^\alpha W}{P_{tx}K}}f_d(x)dx.
\end{equation}

The value of the parameter $p$ lies within an interval $(0, 1)$ and plays a prominent role in the further investigations. It determines the quality of communication. A larger value  of parameter $p$ leads to a better quality of connection and consequently a bigger size of clusters.

%
%
%

\section{Connectivity of the network} \label{Sec:3}

\subsection{Distribution of number of clusters } \label{Sec:NumCl}

The main achievement of this article is that we drive not only the average values of various connectivity characteristics, as in the previous articles (see the Introduction), but also their probability distributions and variations. In this section we investigate such a significant characteristic of the network as distribution of number of clusters. In this section we establish the following formula for the probability of number of clusters $ClustNum$ equaling $r$:
\begin{equation}\label{2121}
P(ClustNum=r)=\binom{n-1}{r-1} p^{n-r}(1-p)^{r-1},
\end{equation}
where parameter $p$ is determined in (\ref{ppppp}).
In other words, number of clusters has binomial distribution.

To prove formula (\ref{2121}) we need to better understand the structure of the network. The idea behind the derivation is to consider connections between vehicles as Bernoulli trials. There are $n$ vehicles in the network establishing no more than $n-1$ connections between each other, thus, we can think about them as $n-1$ independent trials, in which each of them is successful if consecutive vehicles can connect, and unsuccessful otherwise. Clusters are assumed to be formed by a series of several consecutive vehicles, therefore, if there are $r$ clusters in the network, then exactly $r-1$ vehicles cannot establish a connection. From the above we could conclude that the probability $P(ClustNum=r)$ equals the probability that in $n-1$ trials exactly $r-1$ are unsuccessful with the probability of success $p$. This probability is given by well-known formula for binomial distribution, which in our case is  exactly formula (\ref{2121}).

By using known formulas for mean and variance of binomial distribution one can derive that the average number of clusters and their variance are given by the formulas
\begin{equation}\label{55}
\textbf{E}[ClustNum]=1+(n-1)(1-p),
\end{equation}
\begin{equation}
\textbf{Var}[ClustNum]=(n-1)p(1-p).
\end{equation}
Here we consider a disconnected vehicle as a separate cluster. The average number of clusters formed by at least two cars can be calculated by substracting from (\ref{55}) the average number of idle cars derived later in section \ref{idle} (see the formula (\ref{ttt3333})). Therefore, it is given by the following formula:
\begin{equation}
1+(n-1)(1-p)-2(1-p)-(n-2)(1-p)^2.
\end{equation}
\textbf{Remark. } We see from (\ref{55}) that the average number of clusters in the system grows linearly with the number of vehicles with a growth factor $1-p$.

\subsection{Distribution of cluster size}

In this subsection we concentrate on finding probability $P(ClustSize~=~r)$ that size of arbitrary cluster in the network is $r$.
We prove the following formula:

\begin{equation}\label{67}
P(ClustSize=r)=\begin{cases}
                      p^{r-1}(1-p), & \mbox{if } r<n, \\
                      p^{n-1}, & \mbox{if }r=n.
                    \end{cases}
\end{equation}
Unfortunately, we do not know a simple proof. So, we use mathematical apparatus of generating functions, which allows for reformulating the problem in terms of series and use different analytical methods to operate with them.

We denote number of clusters of some vehicle network $N$ by $clust(N)$ and number of clusters having size $r$ by $count(r;N)$. Let us establish the formula
\begin{equation}
P(ClustSize=r)=\sum_{k=1}^n \sum_{N:clust(N)=k}\frac{count(r;N)}{k}  P(N),
\end{equation}
where $P(N)$ is the probability of the network $N$ and the second summation is carried out over all networks $N$ with the restriction $clust(N) = k$.
The network could have from $1$ to $n$ clusters, so $k$ varies from $1$ to $n$, and for each value of the parameter $k$, we consider all networks with $k$ clusters. Finally, for each such case, we sum up the probabilities that a randomly selected cluster has size $r$. Since the network $N$ contains $count(r;N)$ clusters of size $r$, then the probability of choosing a cluster of size $r$ among $k$ clusters is given as $\frac{count(r;N)}{k}$. We should multiply the last probability by $P(N)$ in order to calculate the probability that the cluster of size $k$ is chosen in the network $N$.

We denote networks having $k$ clusters, under condition that $s$ of them have size $r$ by $N_{s,k}$,
therefore, $count(r;N_{s,k})=s$. Since the network has $k$ clusters, $s$ varies from $1$ to $k$. We rewrite the summation over all networks $N$ as the summation over the networks $N_{s,k}$:


\begin{equation}
\label{xxx}
P(ClustSize=r)=
\sum_{k=1}^n \sum_{s=1}^k \sum_{N_{s,k}}\frac{count(r;N_{s,k})}{k}  P(N_{s,k})=
\sum_{k=1}^n \sum_{s=1}^k \sum_{N_{s,k}}\frac{s}{k}  P(N_{s,k}).
\end{equation}
Repeating the steps of derivation of (\ref{2121}) we conclude that
\begin{equation}\label{ffgr}
P(N_{s,k})=(1-p)^{k-1}p^{n-k}.
\end{equation}
From (\ref{xxx}) and (\ref{ffgr}) we derive that
\begin{equation}
\label{ffgr1}
P(ClustSize=r)=\sum_{k=1}^n \sum_{s=1}^k \sum_{N_{s,k}}\frac{s}{k} (1-p)^{k-1}p^{n-k}=
\sum_{k=1}^n (1-p)^{k-1}p^{n-k}\sum_{s=1}^k \frac{s Num(s,k)}{k}.
\end{equation}

Let us establish the formula
\begin{equation}\label{fffg}
Num(s,k)=coeff_{x^n}\binom{k}{s} x^{r s} (x+x^2+x^3+\ldots - x^r)^{k-s},
\end{equation}
where $\binom{k}{s}$ is a binomial coefficient and $coeff_{x^n} P(x)$ is a coefficient of the term $x^n$ in polynomial $P(x)$. The important step is to understand that the value of $Num(s,k)$ equals the number of representations of the integer $n$ as a sum of $k$ summands under condition that exactly $s$ of them equal $r$. This is true, because the clusters are formed by group of consecutive vehicles. Therefore, we can use the formula (\ref{1212dwew1333}) in the appendix identical to the formula (\ref{fffg}) for calculation of $Num(s,k)$.

From (\ref{fffg}) we obtain
\begin{equation}
\label{xxx123}
 \sum_{s=1}^k \frac{s Num(s,k)}{k} = \frac{1}{k} coeff_{x^n} \sum_{s=1}^k \binom{k}{s}s x^{r s} (x+x^2+x^3+\ldots - x^r)^{k-s}= \frac{1}{k} coeff_{x^n} \sum_{s=1}^k \binom{k}{s} s x^{r s} \Big(\frac{x}{1-x} - x^r\Big)^{k-s}.
\end{equation}
The following identity
\begin{equation}\label{ere}
\sum_{s=1}^k \binom{k}{s} s x^{r s} y^{k-s}=k(x^{r}+y)^{k-1}x^r
\end{equation}
holds. We substitute $y=\frac{x}{1-x} - x^r$ into (\ref{ere})
and use (\ref{xxx123}) to derive the following:
\begin{equation}
\label{xxx1234}
\sum_{s=1}^k \frac{s Num(s,k)}{k} \\=
coeff_{x^n}  \Big(x^{r}+\frac{x}{1-x} - x^r\Big)^{k-1}x^r \\=
coeff_{x^n} \Big(\frac{x}{1-x}\Big)^{k-1}x^r.
\end{equation}
Combining equations (\ref{ffgr1}) and (\ref{xxx1234}) we obtain
\begin{multline}
\label{xxx123451}
P(ClustSize=r)=
\sum_{k=1}^n coeff_{x^n} \Big(\frac{x}{1-x}\Big)^{k-1}x^r (1-p)^{k-1}p^{n-k}\\=
p^{n-1} \sum_{k=1}^n coeff_{x^n} \Big(\frac{x}{1-x}\Big)^{k-1}x^r \Big(\frac{1-p}{p}\Big)^{k-1}  =
p^{n-1}\sum_{k=1}^\infty coeff_{x^{n-r}} \Big(\frac{x}{(1-x)}\frac{1-p}{p} \Big)^{k-1}.
\end{multline}
The last equality in (\ref{xxx123451}) is true because of the identity
$$
p^{n-1}\sum_{k=n+1}^\infty coeff_{x^{n-r}} \Big(\frac{x}{(1-x)}\frac{1-p}{p} \Big)^{k-1}=0.
$$
The last identity is valid, because the series $\sum_{k=n+1}^\infty  \Big(\frac{x}{(1-x)}\frac{1-p}{p} \Big)^{k-1}$ has nonzero coefficients only of $x^i,i\geq n$. Therefore, the coefficient of $x^{n-r}$ equals zero.
Applying the formula of the sum of geometric progression to (\ref{xxx123451}) we finally derive
\begin{multline}
\label{xxx12345}
P(ClustSize=r)=
p^{n-1}coeff_{x^{n-r}} \sum_{k=1}^\infty  \Big(\frac{x}{(1-x)}\frac{1-p}{p}\Big)^{k-1} =
p^{n-1}coeff_{x^{n-r}}  \frac{1}{1-\frac{x}{(1-x)}\frac{1-p}{p}}=
p^{n-1}coeff_{x^{n-r}}  \frac{1-x}{1-x(1+\frac{1-p}{p})}\\=
p^{n-1}coeff_{x^{n-r}} (1-x) \sum_{k=0}^\infty x^k p^{-k} =
\begin{cases}
        p^{n-1} \Big( p^{-n+r} -p^{-n+r+1}\Big), & \mbox{if } r<n, \\
        p^{n-1}, & \mbox{if } r=n
\end{cases}
=\begin{cases}
        p^{r-1}(1-p), & \mbox{if } r<n, \\
        p^{n-1}, & \mbox{if } r=n.
\end{cases}
\end{multline}
Formula (\ref{67}) is proven.

Now we are ready to calculate the average value and the variance of the distribution (\ref{67}). Firstly, we find the first two moments by the formulas
\begin{equation}
M_1=np^{n-1}+(1-p)\sum_{\alpha=1}^{n-1} \alpha p^{\alpha-1},
\end{equation}
\begin{equation}
M_2=n^2p^{n-1}+(1-p)\sum_{\alpha=1}^{n-1} \alpha^2 p^{\alpha-1}.
\end{equation}
By using the identities
\begin{equation}
\sum_{\alpha=1}^{n-1} \alpha p^{\alpha-1}=\frac{1+((n-1)p-n)p^{n-1}}{(1-p)^2},
\end{equation}
\begin{equation}
  \sum_{\alpha=1}^{n-1} \alpha^2 p^{\alpha-1}=\frac{1+p-((n-1)^2p^2+(-2n^2+2n+1)p+n^2)p^{n-1}}{(1-p)^3}
\end{equation}
we derive
\begin{equation}\label{156}
M_1=\frac{1-p^n}{1-p},
\end{equation}
\begin{equation}\label{157}
M_2=\frac{2p^{n+1}n-2p^nn-p^{n+1}-p^n+p+1}{(1-p)^2}.
\end{equation}
From formulas (\ref{156}) and (\ref{157}) we finally obtain
\begin{equation}\label{1156}
\textbf{E}[ClustSize]=M_1=\frac{1-p^n}{1-p},
\end{equation}
\begin{equation}\label{1157}
\textbf{Var}[ClustSize]=M_2-M_1^2\\=\frac{2p^{n+1}n-2p^n n-p^{2n}-p^{n+1}+p^n+p}{(1-p)^2}.
\end{equation}

\textbf{Remark. } It follows from the formulas (\ref{1156}) and (\ref{1157}) that the \textbf{average cluster size approximately equals ${1/(1-p)}$ (constant!) and its variance approximately equals ${p/(1-p)^2}$}, since $p^n\rightarrow 0, n\rightarrow\infty$.

\subsection{Distribution of size of the biggest cluster}

In the previous section we explored the question about distribution of cluster size. It was established that the average cluster size is
approximately $1/(1-p)$, therefore, being a constant. However, it is natural to investigate this problem further and understand how the biggest cluster size deviates from the average.


We denote the probability that the biggest cluster of the network is formed by exactly $r$ vehicles by $P(BiggestClust=r)$. Let us introduce the function $g(k)$ by the formula
\begin{equation}\label{sgsgdh}
g(k)=\sum_{m=1}^{\lfloor n/(k+1) \rfloor } (-1)^m p^{mk}(1-p)^{m-1} \binom{n-1-mk}{m-1}+
\sum_{m=0}^{\lfloor (n-1)/(k+1) \rfloor } (-1)^m p^{mk}(1-p)^{m}\binom{n-1-mk}{m},
\end{equation}
where $\lfloor x \rfloor$ means rounding down. We establish the following formula:
\begin{equation}\label{longest_run}
P(BiggestClust=r)=\begin{cases}
                      p^{n-1}, & \mbox{if }r=n,\\
                      g(r)-g(r-1), & \mbox{if } 1<r<n, \\
                      (1-p)^{n-1}, & \mbox{if }r=1.
                    \end{cases}
\end{equation}
The cases $r=n$ and $r=1$ are trivial, because under these assumptions all vehicles are connected or disconnected respectively. Thus, we need to prove the formula (\ref{longest_run}) only in the case $1<r<n$.
We denote the length of the longest run in the sequence of the $n-1$ Bernoulli trials by $L_{n-1}$. The following equation is proven in \cite{lr3}:
\begin{equation}\label{sdfs}
P(L_{n-1}\leq r-1)=g(r),
\end{equation}
where $g(k)$ is determined in (\ref{sgsgdh}). The number of vehicles in the network is $n$, consequently, there are $n-1$ connections between them and we can consider them as a sequence of Bernoulli trials as before. The probability $P(BiggestClust~=~r)$ equals the probability that the network has the longest run equaling $r-1$ (number of connections is less than the number of cars by 1). From the above we derive
\begin{equation}\label{sdfs11}
P(BiggestClust=r)=P(L_{n-1}=r-1)=P(L_{n-1}\leq r-1)-P(L_{n-1}\leq r-2)=g(r)-g(r-1).
\end{equation}
Formula (\ref{longest_run}) is proven.

Despite the fact that the formula (\ref{longest_run}) can be used for the calculation of the probabilities $P(BiggestClust=r)$, it is overly complicated to predict behaviour of the network with growing number of vehicles.
Fortunately, the asymptotic of the longest successful run with growing $n$ is already found (see, for instance, \cite{LR,LR2}),
in which the expected value and variance of the longest run are derived for sufficiently large values of $n$. Taking into account that there are $n-1$ connections in our network, the fact that the number of connections is less that number of vehicles in a cluster by 1, and the formulas from above mentioned articles we conclude that the expected value and variation of the biggest cluster of our network are
\begin{equation}\label{approx}
\textbf{E}[BiggestClust]\approx \log_{1/p}\{(n-1)(1-p)\}+\gamma/\ln(1/p)+0.5,
\end{equation}
\begin{equation}
\textbf{Var}[BiggestClust]\approx \pi^2/\ln^2(1/p)+\frac{1}{12},
\end{equation}
where $\gamma=0.577\ldots$ is Euler constant.

\textbf{Remark. } From the previous derivations we know that the size of average cluster is $\approx 1/(1-p)$, however, from (\ref{approx}) we conclude that the size of the biggest cluster grows logarithmically with the number of vehicles in the network.

\subsection{Distribution of idle cars} \label{idle}

We call car \textit{idle} if it fails to establish a connection with any of its neighbours.
Unfortunately, there is no simple formula for calculating the probability $P(IdleCars=r)$ that the network has exactly $r$ \textit{idle} vehicles. One can prove by using method of generating functions that the probability $P(IdleCars=r)$ can be calculated by the formula
\begin{equation}\label{ttuu}
 P(IdleCars=r)= \frac{p^{n+1}}{(1-p)^2(n-r)!}\frac{d^{n-r}}{dx^{n-r}}\Big\{\frac{(1-x)^{r+1} }{((1-x)\frac{p}{1-p}-x^2)^{r+1}}\Big\}\Big|_{x=0},
\end{equation}
where $k!$ is a factorial and $\frac{d^k}{dx^k}f(x)\Big|_{x=0}$ is $k$-th derivative of function $f(x)$ calculated at the point $x=0$.

Despite the fact that the formula of distribution of \textit{idle} vehicles is quite complicated, we can derive simple expression for the expected number of \textit{idle} vehicles. Let us denote the random variable that equals 1 it $k$-th vehicle is idle and 0 otherwise by $\xi_k$.
Let us prove that the expected value of  $\xi_k$ can be determined by this formula
\begin{equation}
\label{ttt}
\textbf{E}[\xi_k]=\begin{cases}
      (1-p)^2, & \mbox{if } 1<k<n, \\
      1-p, & \mbox{if } k=1 \mbox{ or } k=n.
    \end{cases}
\end{equation}
It is enough to prove the formula (\ref{ttt}) in the case $1<k<n$, since derivations in other cases are similar. If $1<k<n$, then the vehicle has exactly two neighbours. Since the probability of disconnection equals $(1-p)$, the probability of car being \textit{idle} equals $(1-p)^2$. Finally,
\begin{equation}
\label{ttt2212}
\textbf{E}[\xi_k]=1\times(1-p)^2+0\times(1-(1-p)^2)=(1-p)^2.
\end{equation}
Formula (\ref{ttt}) is proven.

Therefore, the average number of \textit{idle} cars is
\begin{equation}\label{ttt3333}
\textbf{E}[IdleCars]=\textbf{E}\Big[\sum_{k=1}^n\xi_k\Big]=\sum_{k=1}^n\textbf{E}[\xi_k]=2(1-p)+\sum_{k=2}^{n-1} (1-p)^2=2(1-p)+(n-2)(1-p)^2.
\end{equation}

Let us calculate the variance of distribution of number of \textit{idle} vehicles. It is given as follows:
\begin{equation}\label{ttt1111}
\textbf{Var}[IdleCars]=\textbf{E}\Big[\Big(\sum_{k=1}^n\xi_k\Big)^2\Big]-\textbf{E}[IdleCars]^2=
\sum_{k=1}^n \textbf{E} [\xi_k^2]+2\sum_{k=1}^{n-1} \sum_{j=k+1}^n \textbf{E}[\xi_k\xi_j]-\textbf{E}[IdleCars]^2.
\end{equation}
If two vehicles are not neighbours then the variables $\xi_k$ and $\xi_j$ are independent and, therefore, $\textbf{E}[\xi_k\xi_j]=\textbf{E}[\xi_k]\textbf{E}[\xi_j]$, otherwise $\textbf{E}[\xi_k\xi_j]$ can be derived by analogy with the derivation of (\ref{ttt}). Thus, the following formula
\begin{equation}
\label{ttt111}
\textbf{E}[\xi_k\xi_j]=\begin{cases}
      \textbf{E}[\xi_k]\textbf{E}[\xi_j], & \mbox{if } j>k+1, \\
      (1-p)^3, & \mbox{if } j=k+1 \mbox{ and } 1<k<n-1,\\
      (1-p)^2, & \mbox{if } j=k+1, k=1 \mbox{ or } k=n-1.
    \end{cases}
\end{equation}
is valid for $j> k$. Using (\ref{ttt111}), we derive
\begin{multline}\label{tttrr}
\sum_{k=1}^{n-1} \sum_{j=k+1}^n \textbf{E}[\xi_k\xi_j]=
\sum_{k=1}^{n-1} \sum_{j=k+2}^{n} \textbf{E}[\xi_k]\textbf{E}[\xi_j]+
\sum_{k=2}^{n-2} \textbf{E}[\xi_k\xi_{k+1}]+
\textbf{E}[\xi_1\xi_2]+\textbf{E}[\xi_{n-1}\xi_n]\\=
\sum_{k=1}^{n-1} \sum_{j=k+2}^{n} \textbf{E}[\xi_k]\textbf{E}[\xi_j]+(n-3)(1-p)^3+2(1-p)^2.
\end{multline}
Consequently, by using (\ref{ttt1111}) and (\ref{tttrr}) and taking into account that $\xi_k^2=\xi_k$ we obtain
\begin{multline}\label{ttt11111}
\textbf{Var}[IdleCars]=
\sum_{k=1}^n \textbf{E} [\xi_k^2]+2\sum_{k=1}^{n-1} \sum_{j=k+1}^n\textbf{E}[\xi_k\xi_j]-
\Big(\textbf{E}\Big[\sum_{k=1}^n\xi_k\Big]\Big)^2=
\sum_{k=1}^n \textbf{E} [\xi_k^2]+2\sum_{k=1}^{n-1} \sum_{j=k+2}^{n} \textbf{E}[\xi_k]\textbf{E}[\xi_j]\\+2(n-3)(1-p)^3+4(1-p)^2-
\sum_{k=1}^n \textbf{E} [\xi_k]^2-2\sum_{k=1}^{n-1} \sum_{j=k+1}^n \textbf{E}[\xi_k]\textbf{E}[\xi_j]
\\=\sum_{k=1}^n \textbf{E} [\xi_k]-\sum_{k=1}^n \textbf{E} [\xi_k]^2
-2\sum_{k=1}^{n-1}\textbf{E}[\xi_k]\textbf{E}[\xi_{k+1}]+2(n-3)(1-p)^3+4(1-p)^2.
\end{multline}
We derive the explicit formulas for every sum in (\ref{ttt11111}). Equality (\ref{ttt3333}) gives us
\begin{equation}
\label{yyyy}
\sum_{k=1}^n \textbf{E} [\xi_k]=2(1-p)+(n-2)(1-p)^2.
\end{equation}
By analogy with derivation of (\ref{ttt3333}) we can deduce that
\begin{equation}
\label{yyyy11}
\sum_{k=1}^n \textbf{E} [\xi_k]^2=2(1-p)^2+(n-2)(1-p)^4.
\end{equation}
The last sum in (\ref{ttt11111}) can be derived using (\ref{ttt}) as follows:
\begin{equation}\label{uuuurrr}
\sum_{k=1}^{n-1}\textbf{E}[\xi_k]\textbf{E}[\xi_{k+1}]=
\sum_{k=2}^{n-2}\textbf{E}[\xi_k]\textbf{E}[\xi_{k+1}]+
\textbf{E}[\xi_1]\textbf{E}[\xi_2]+\textbf{E}[\xi_{n-1}]\textbf{E}[\xi_n]=(n-3)(1-p)^4+2(1-p)^3.
\end{equation}

Combining formulas (\ref{ttt11111})--(\ref{uuuurrr}) we obtain
\begin{multline}\label{uuuu}
\textbf{Var}[IdleCars]=
2(1-p)+(n-2)(1-p)^2-2(1-p)^2-(n-2)(1-p)^4-2(n-3)(1-p)^4-4(1-p)^3\\+2(n-3)(1-p)^3+4(1-p)^2=-3np^4 + 10np^3 - 11np^2 + 4np + 8p^4 - 22p^3 + 18p^2 - 4p.
\end{multline}

\textbf{Remark. } Formula (\ref{ttt3333}) leads us to an important conclusion that in average there is a constant fraction $\approx (1-p)^2$ of \textit{idle} vehicles in the network.

\section{Simulations} \label{Sec:4}

We use the model described by the equations (\ref{11144})--(\ref{ppppp}) and
make simulations for $n=10$ and $n=20$ vehicles on the road assuming exponential distribution of the intervehicle distance with probability density function $\rho e^{-\rho x}$, where $\rho$ is a vehicle density.
We consider two scenarios. In the first case, the density is low $\rho = 0.01$ (10 cars per kilometer) which is the case for rural traffic, in the second case $\rho = 0.05$ (50 cars per kilometer), which corresponds to urban traffic.
We use the following values of the parameters taken from \cite{ChModel} (see their descriptions in section \ref{Sec:1}):
\begin{center}
$G_T = G_R = 1$, $f_c=5.9$ GHz, $\alpha$ = 2.5, \\
$T_0 =300^\circ$ K, $B = 10$ MHz, $\Psi=10$ dB.
\end{center}
We use the transmit power value $P_{tx} = 4$ dBm. This value corresponds to the message transmission over short distances up to 99 meters \cite{ChModel2}.

Graphs of all network properties are plotted in the cases \textbf{a)} $ n = 10 $, $ \rho = 0.01 $, \textbf{b)} $ n = 10 ,  \rho = 0.05 $, \textbf{c)} $ n = 20 ,  \rho = 0.01 $, and \textbf{d)}~$ n = 20 ,  \rho = 0.05 $. In the case $\rho = 0.01 $, the probability of connection between two consecutive cars is $p = 0.5576$, in the case $ \rho = 0.01 $ the probability of connection is $ p = 0.9525 $ (it does not depend on the number of cars within the model). To verify the obtained formulas we generate sequence of vehicles randomly 100000 times and then calculate network connectivity distributions from these sample values.
The graphics of the simulations are identical to the results obtained by the formulas, which confirms their correctness.
%

On Figure 1 the graphics of distribution (\ref{2121}) of number of clusters are presented as well as simulation results for scenarios \textbf{a}, \textbf{b}, \textbf{c}, and \textbf{d}.
As one can see from Figure 1 in cases \textbf{a} and \textbf{c}, the connection is less stable ($ p = 0.5576 $), therefore, the network on average has sufficiently large number of clusters 4.9812 and 9.4048, respectively. In cases \textbf{b} and \textbf{d}, the connection occurs with a much higher probability $ p = 0.9525 $, therefore, in these cases the network tends to be fully connected, which is expressed in a high probability of having 1--3 clusters. In these cases average cluster sizes are significantly lower: 1.4278 and 1.9031, respectively.
The fact that the distributions on Figure 1 are close to normal is not surprising, since, as well known, the binomial distribution in the case $ n \to \infty $ tends to normal.

\begin{figure}[h]
    \centering 
\begin{subfigure}{0.48\textwidth}
  \includegraphics[width=1\textwidth]{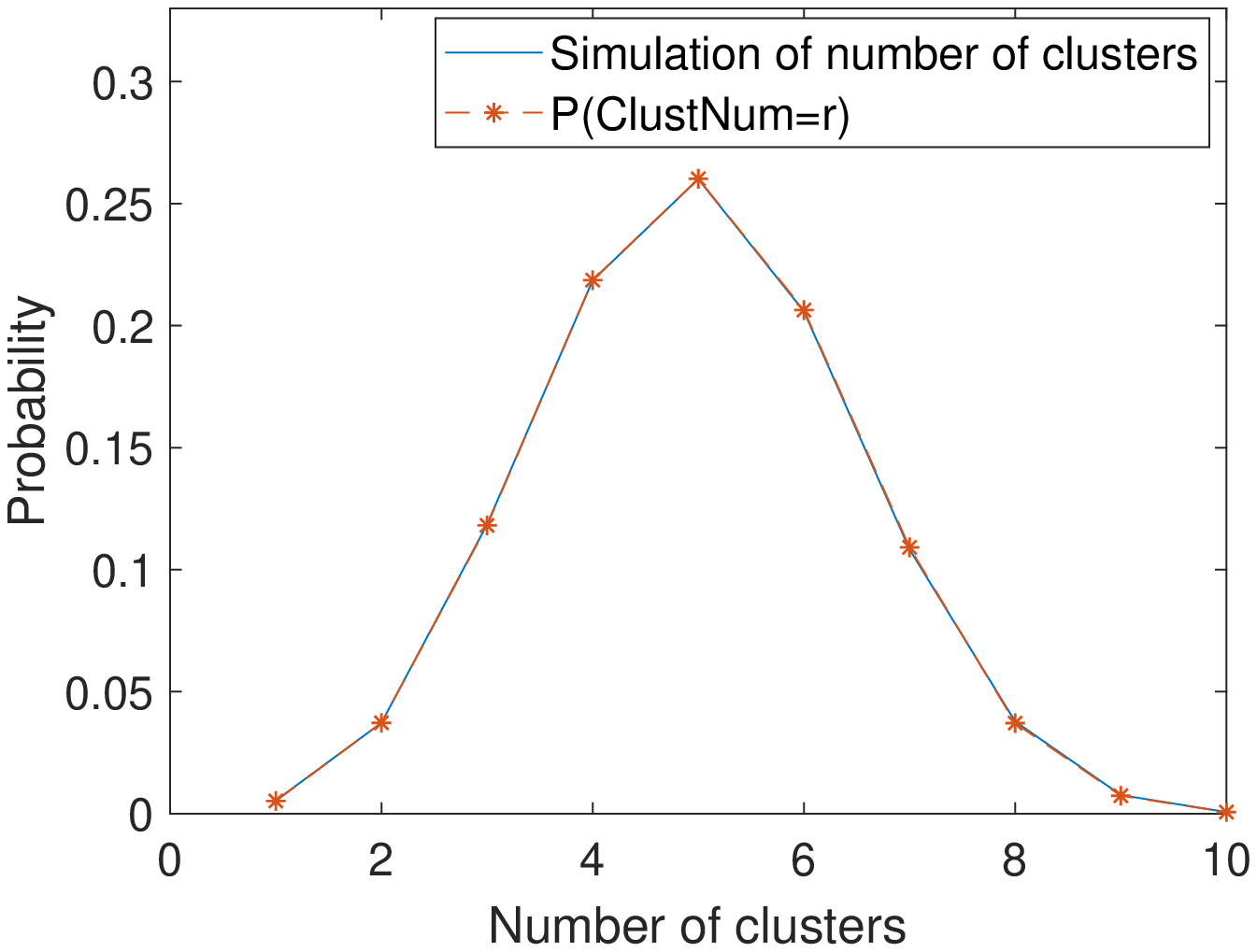}
  \caption{$n=10$,$\rho=0.01$}
  \label{fig:1}
\end{subfigure}\hfil 
\begin{subfigure}{0.48\textwidth}
  \includegraphics[width=1\textwidth]{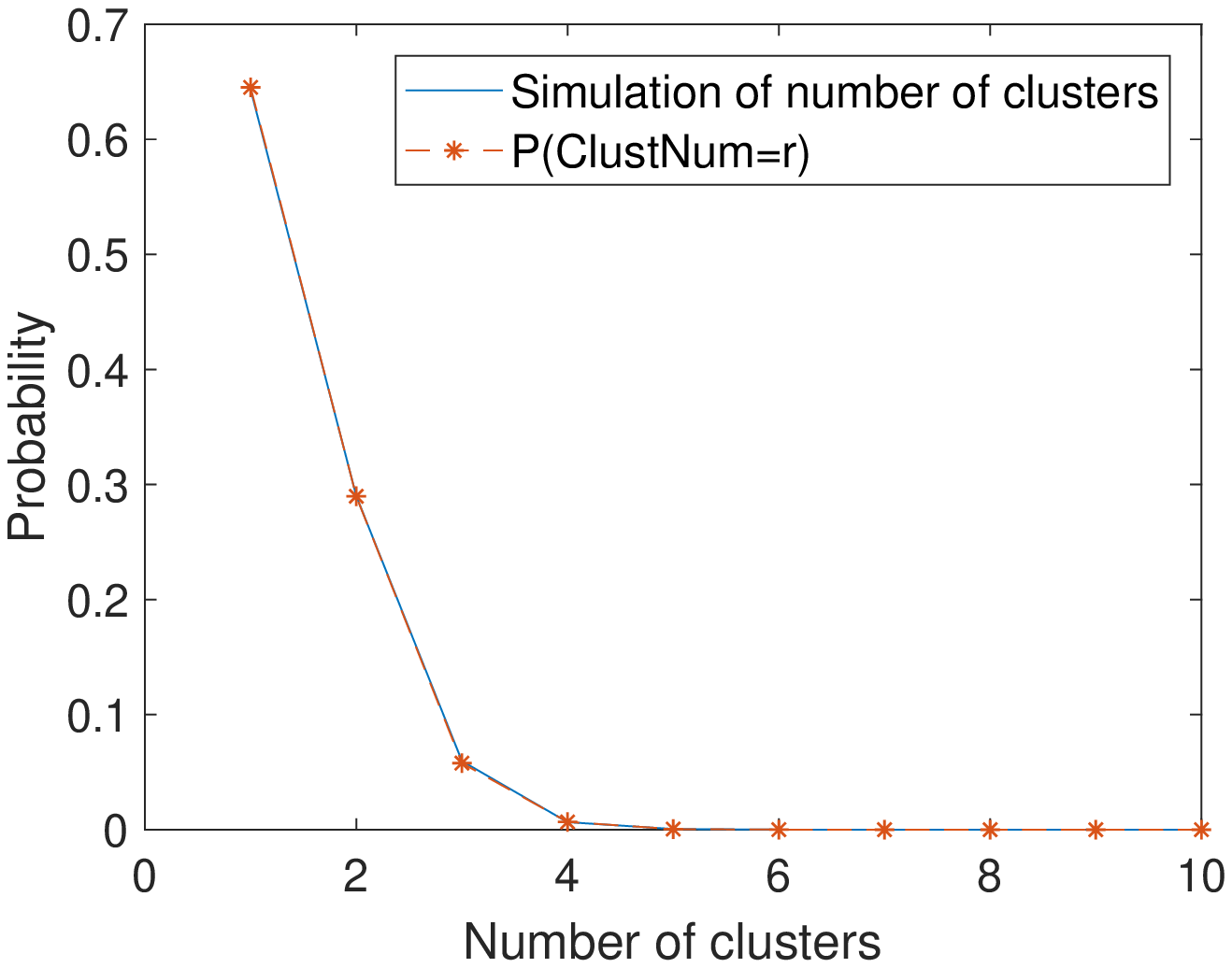}
  \caption{$n=10$,$\rho=0.05$}
  \label{fig:2}
\end{subfigure}

\medskip
\begin{subfigure}{0.48\textwidth}
  \includegraphics[width=1\textwidth]{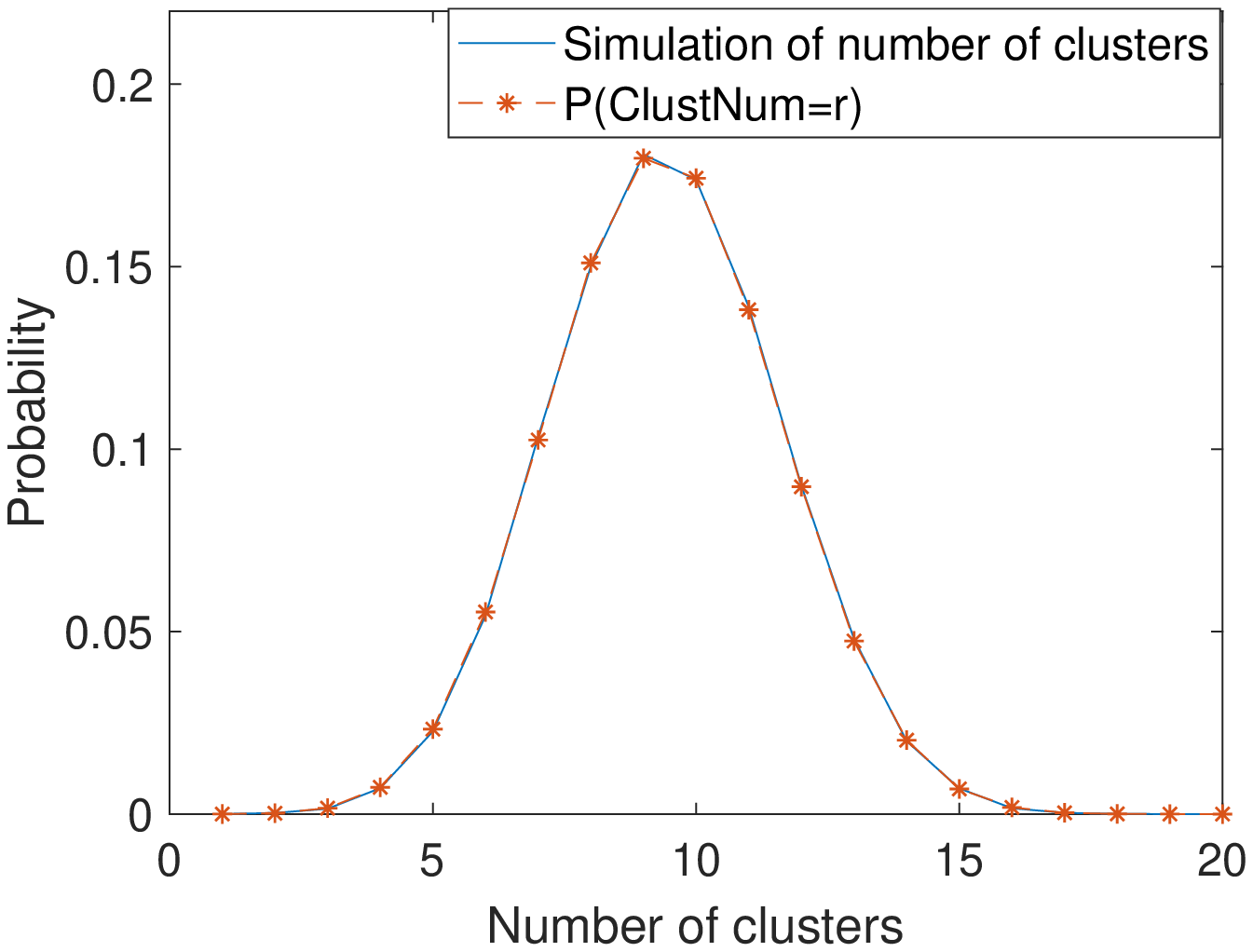}
  \caption{$n=20$,$\rho=0.01$}
  \label{fig:4}
\end{subfigure}\hfil 
\begin{subfigure}{0.48\textwidth}
  \includegraphics[width=1\textwidth]{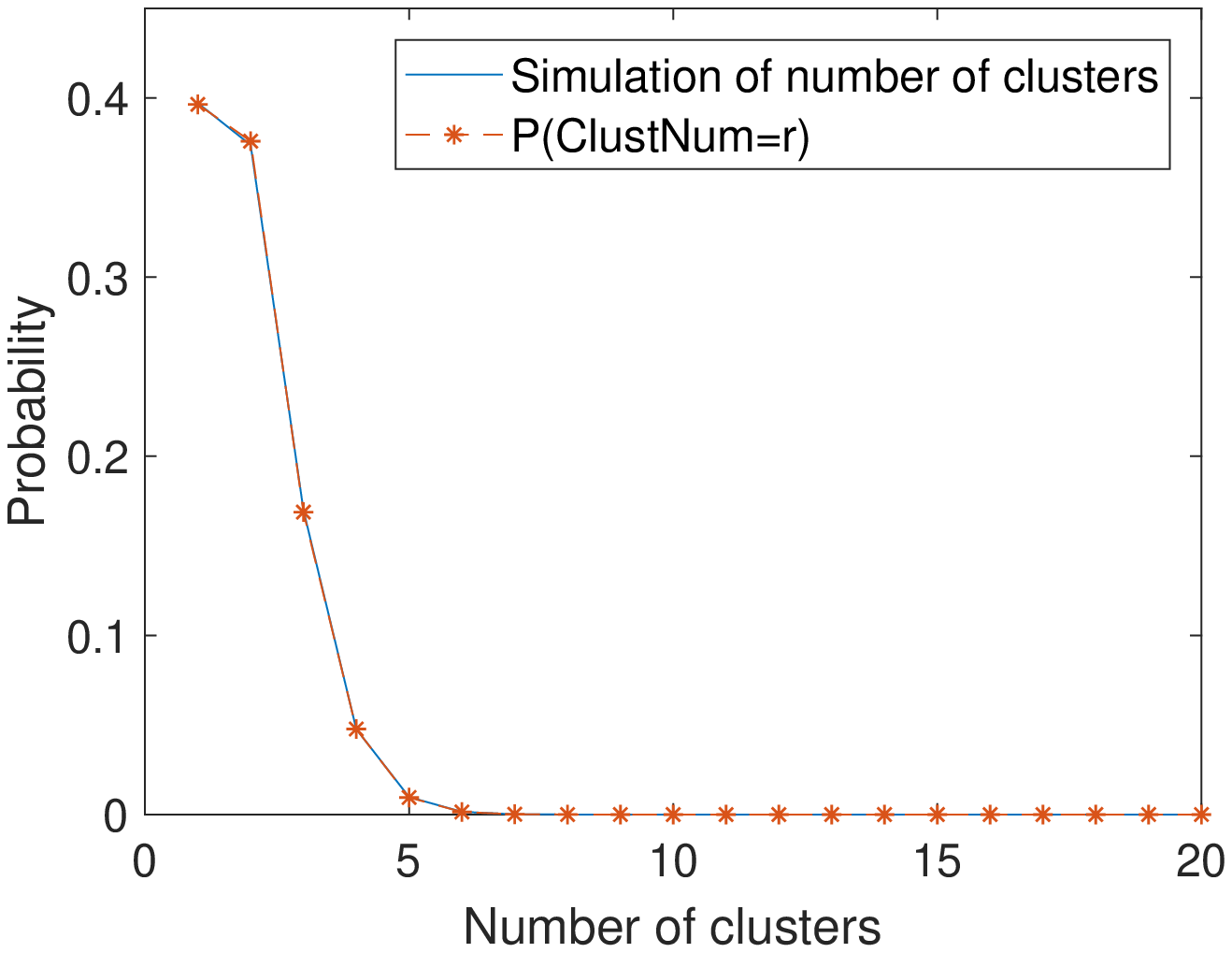}
  \caption{$n=20$,$\rho=0.05$}
  \label{fig:5}
\end{subfigure}
\caption{Probability distribution of number of clusters in the network for different $n$ and $\rho$ }
\label{fig:images}
\end{figure}

On Figure 2 the graphs of the cluster size distribution are presented. The results are obtained by the formula (\ref{67}) and compared to the simulation results.
The same logic as in the previous paragraph leads us to the conclusion that the cluster size in cases \textbf{a} and \textbf{c} should be smaller than in cases \textbf{b} and \textbf{d}. This hypothesis is confirmed, because the average cluster sizes in these cases are 2.2541 and 2.2606, respectively, versus 8.1109 and 13.0948 in cases \textbf{b} and \textbf{d}. The decrease in graphs in cases \textbf{a} and \textbf{c} is explained by the fact that in these cases the connection is less stable, therefore, the probability of a cluster having a size $i$ decreases with increasing of the parameter $ i $.
Graphs \textbf{b} and \textbf{d} have a pronounced peak, since in the case of stable connection, the network tends to be fully connected, therefore, a cluster having size $ n $ has a maximum probability.


\begin{figure}[h]
    \centering 
\begin{subfigure}{0.48\textwidth}
  \includegraphics[width=1\textwidth]{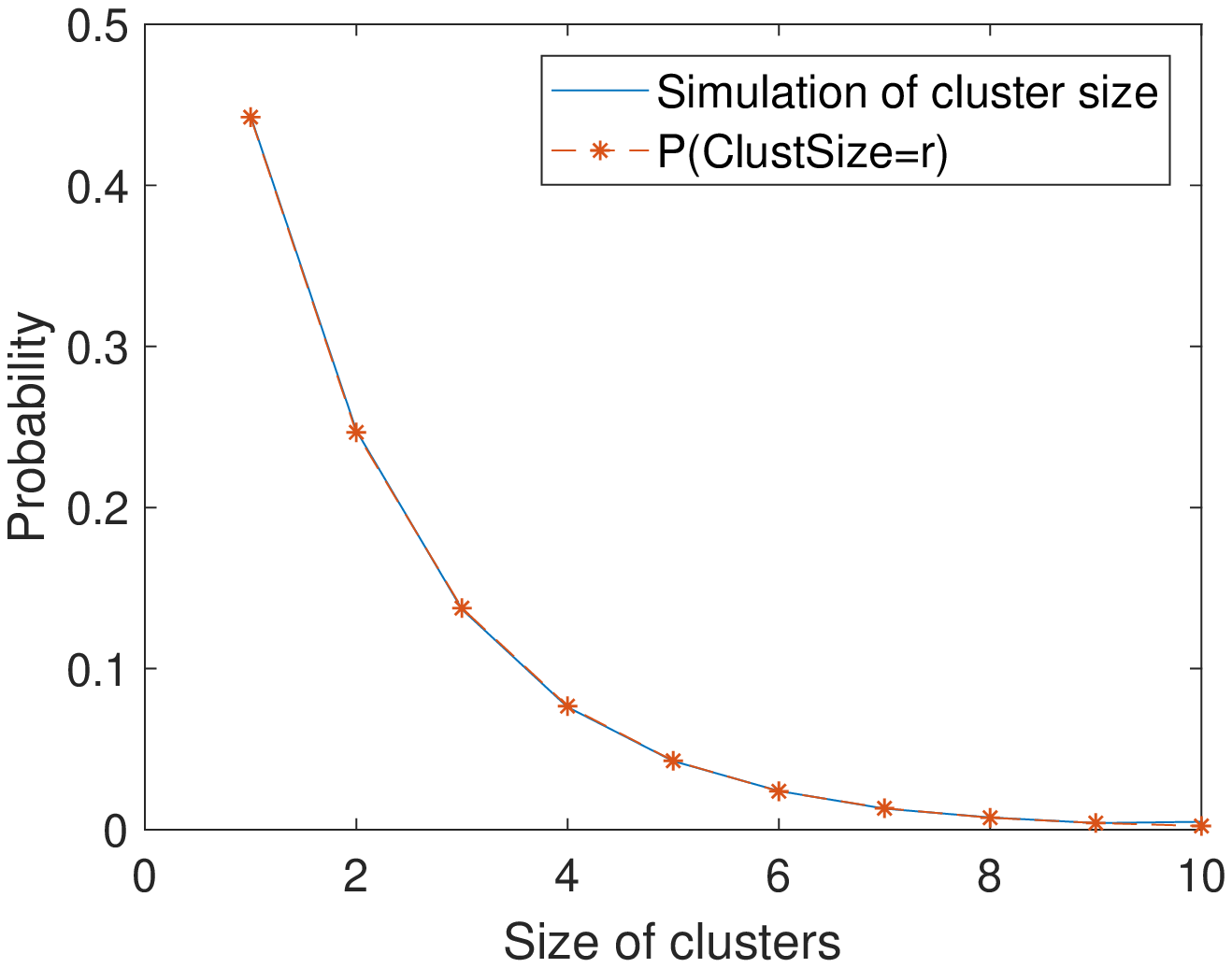}
  \caption{$n=10$,$\rho=0.01$}
  \label{fig:1}
\end{subfigure}\hfil 
\begin{subfigure}{0.48\textwidth}
  \includegraphics[width=1\textwidth]{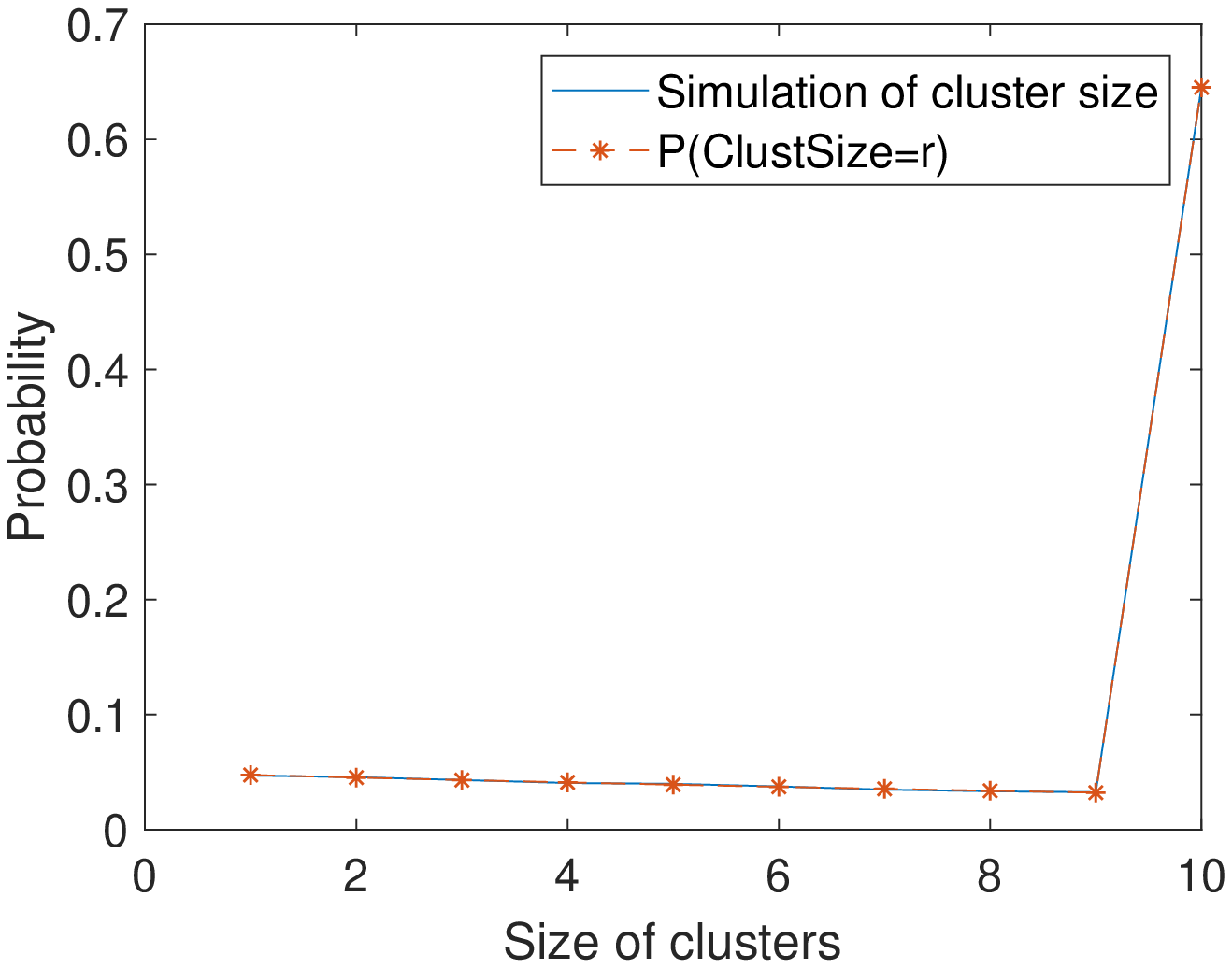}
  \caption{$n=10$,$\rho=0.05$}
  \label{fig:2}
\end{subfigure}

\medskip
\begin{subfigure}{0.48\textwidth}
  \includegraphics[width=1\textwidth]{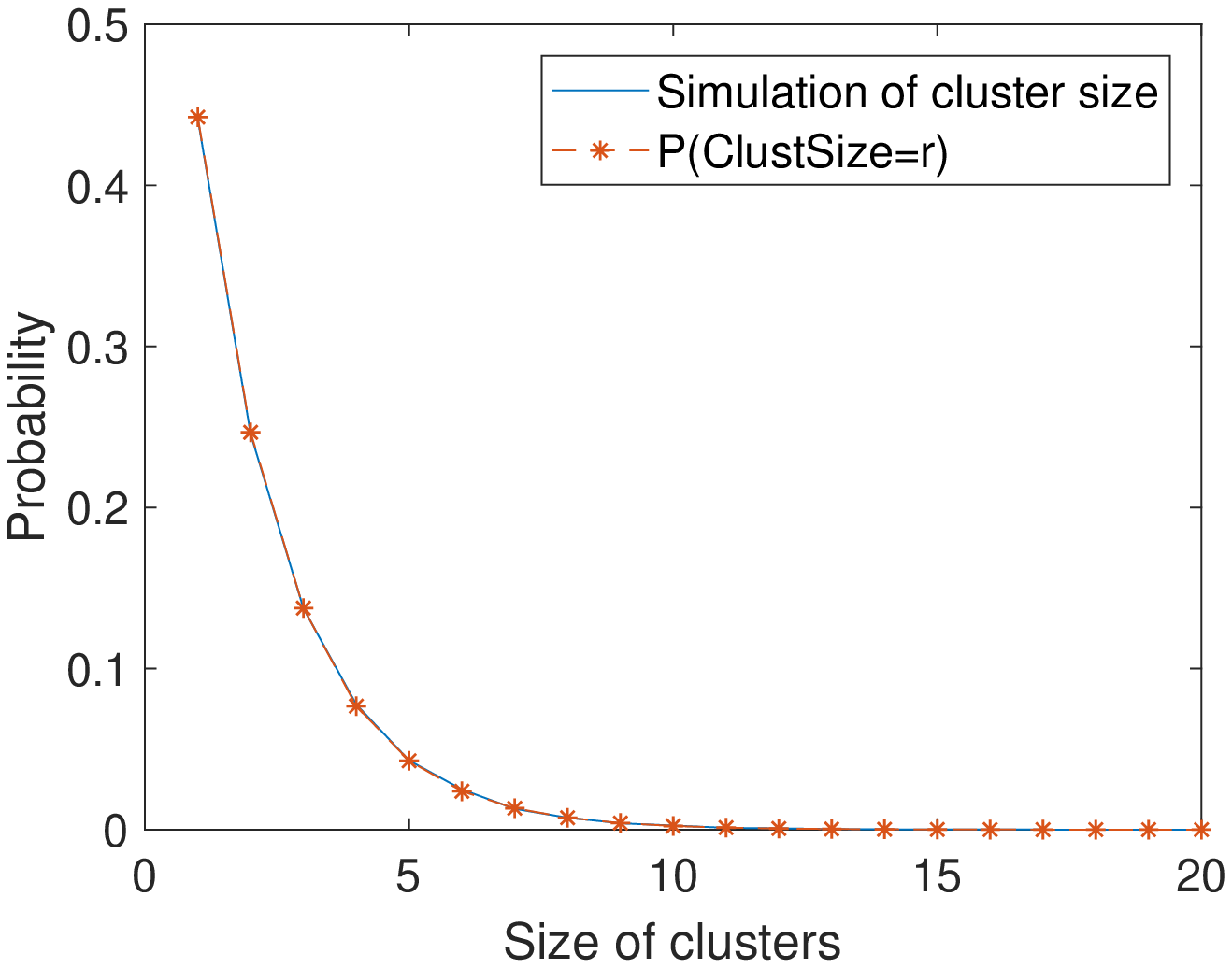}
  \caption{$n=20$,$\rho=0.01$}
  \label{fig:4}
\end{subfigure}\hfil 
\begin{subfigure}{0.48\textwidth}
  \includegraphics[width=1\textwidth]{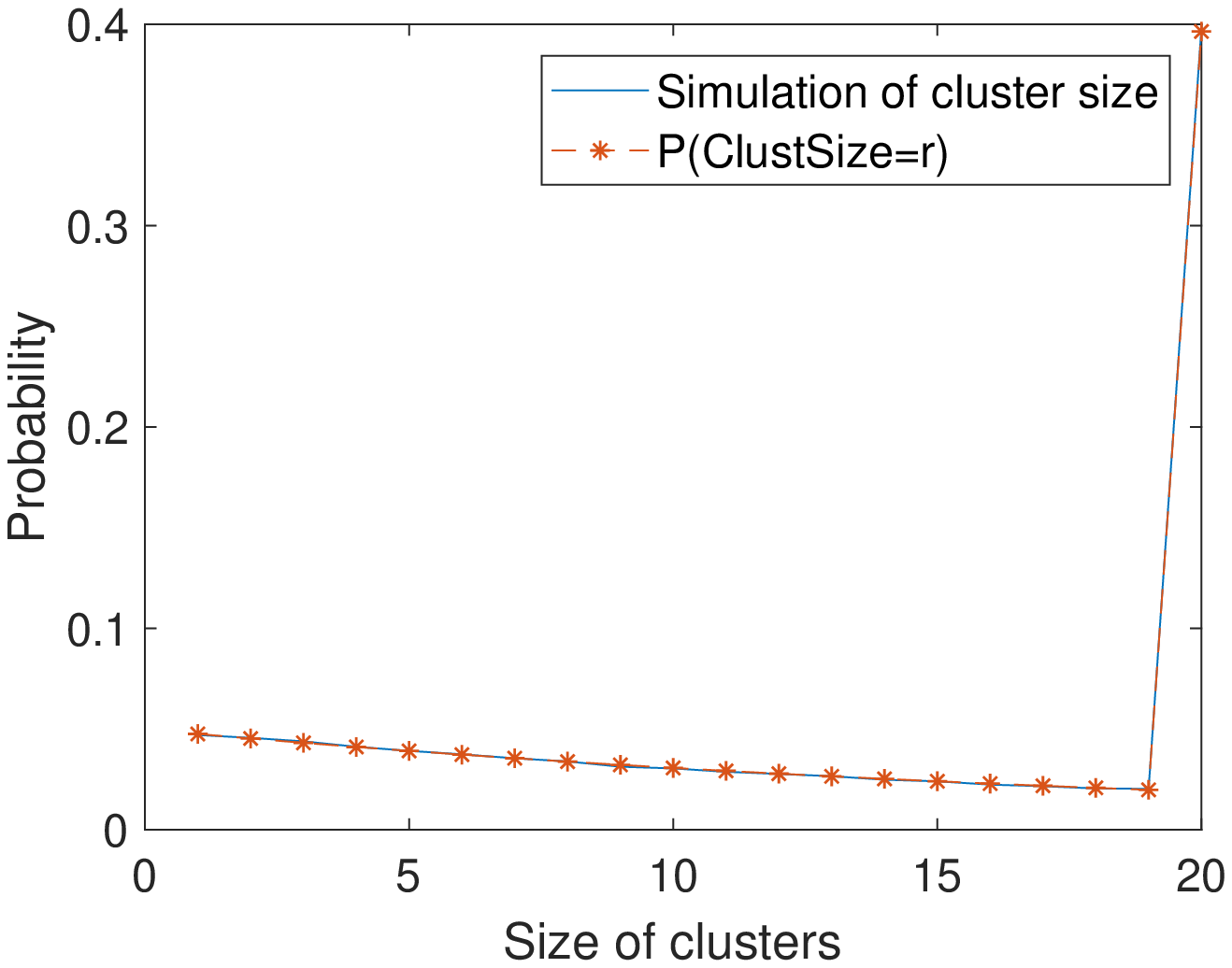}
  \caption{$n=20$,$\rho=0.05$}
  \label{fig:5}
\end{subfigure}
\caption{Probability distribution of size of a cluster for different $n$ and $\rho$ }
\label{fig:images}
\end{figure}

 The graphs of the biggest cluster size distribution obtained by the formula (\ref{longest_run}) and simulation results are presented on Figure 3. By analogy with the explanation of the behavior of cluster size distribution graphs, the behavior of the maximum cluster size distribution can be explained. The average values in cases \textbf{a}, \textbf{b}, \textbf{c}, and \textbf{d} equal 4.0425, 8.9048, 5.2253, and 16.0203, respectively.



\begin{figure}[h]
    \centering 
\begin{subfigure}{0.48\textwidth}
  \includegraphics[width=1\textwidth]{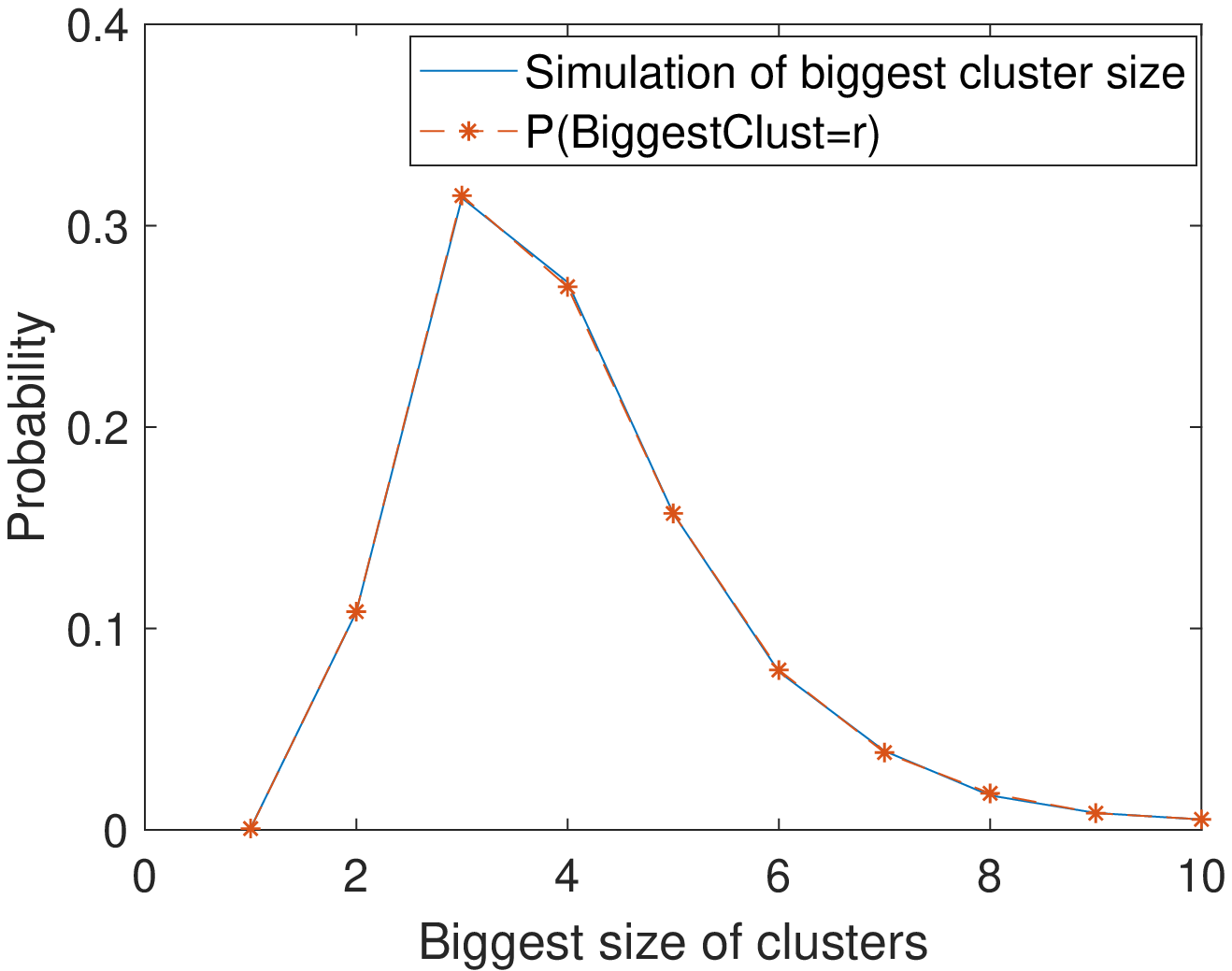}
  \caption{$n=10$,$\rho=0.01$}
  \label{fig:1}
\end{subfigure}\hfil 
\begin{subfigure}{0.48\textwidth}
  \includegraphics[width=1\textwidth]{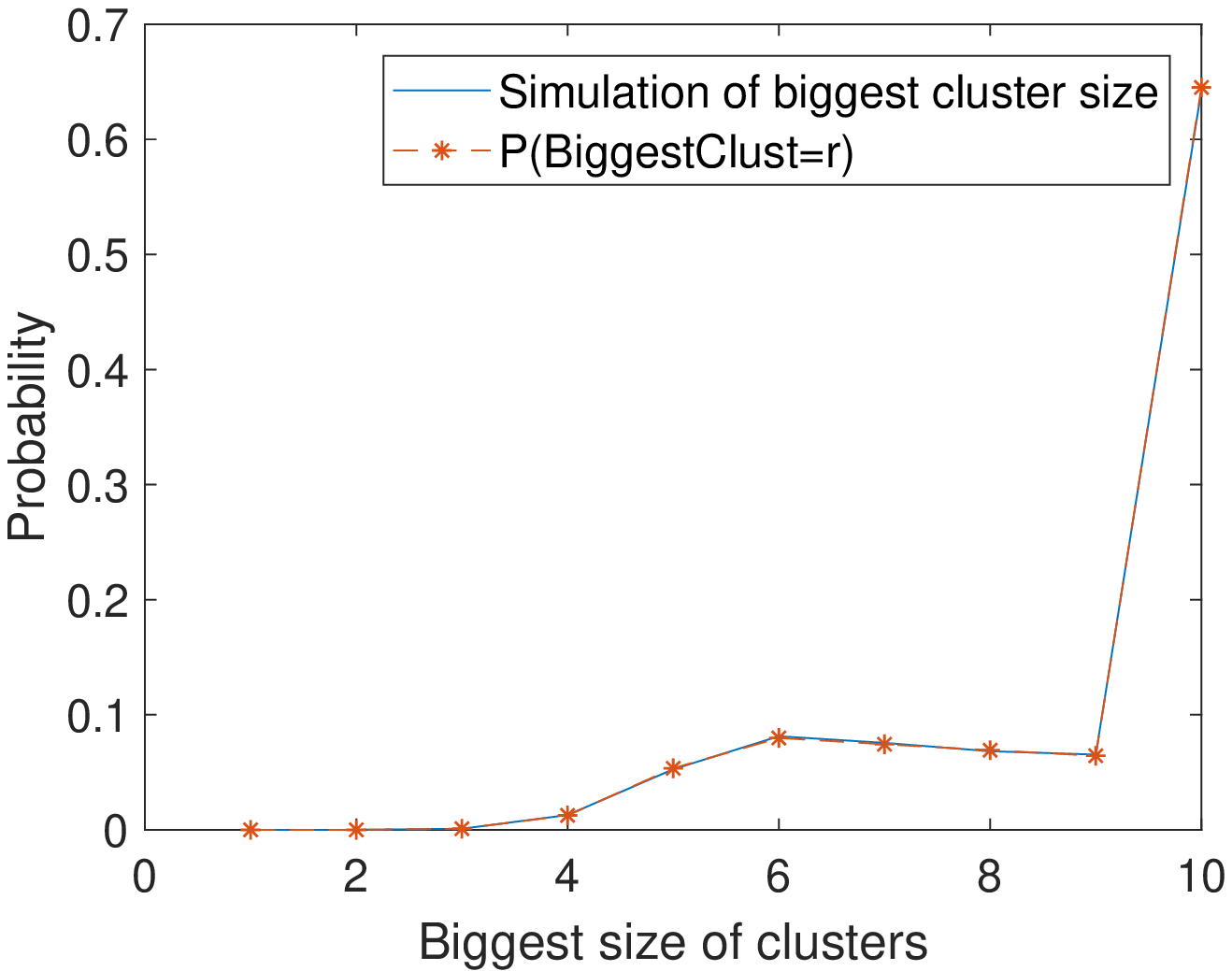}
  \caption{$n=10$,$\rho=0.05$}
  \label{fig:2}
\end{subfigure}

\medskip
\begin{subfigure}{0.48\textwidth}
  \includegraphics[width=1\textwidth]{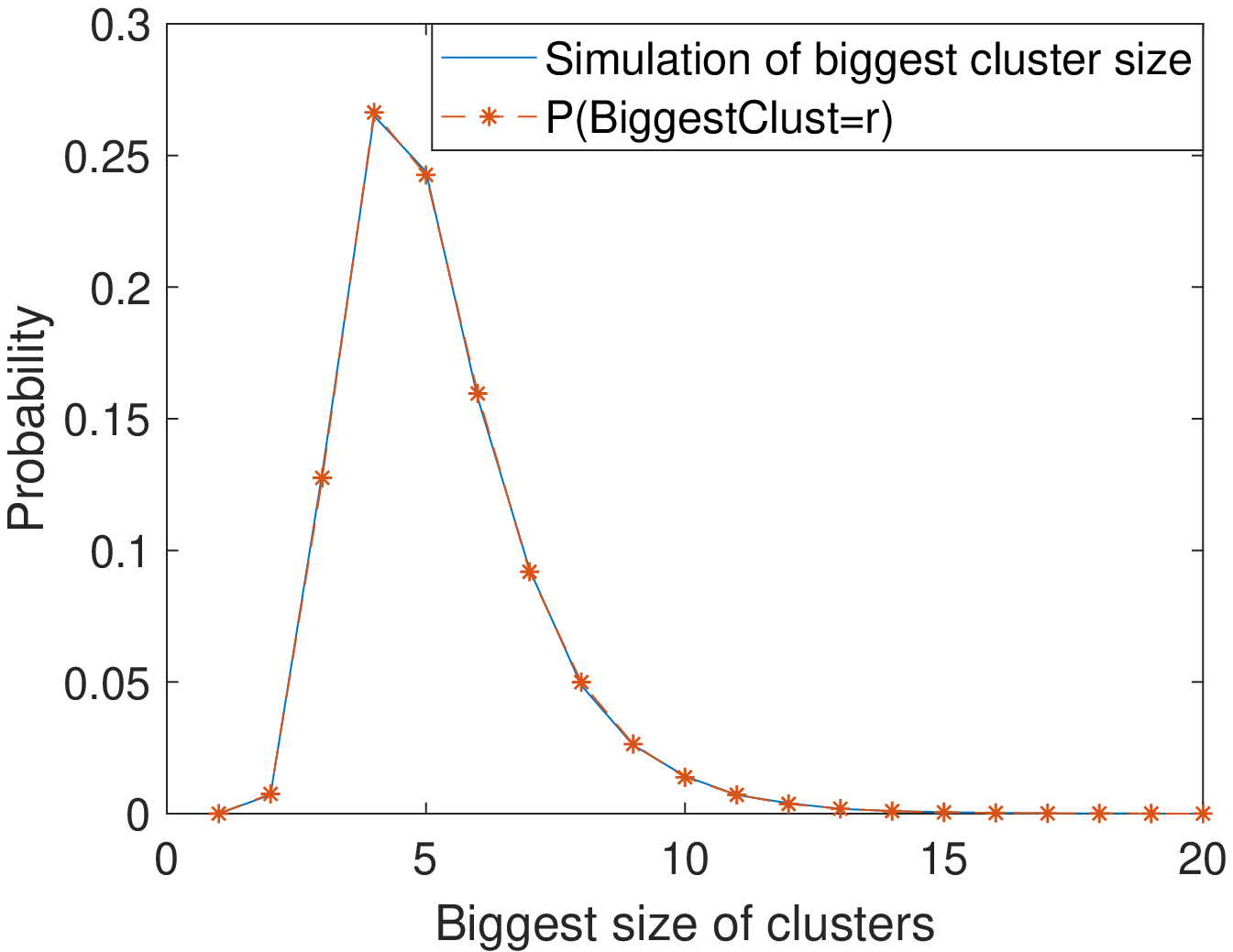}
  \caption{$n=20$,$\rho=0.01$}
  \label{fig:4}
\end{subfigure}\hfil 
\begin{subfigure}{0.48\textwidth}
  \includegraphics[width=1\textwidth]{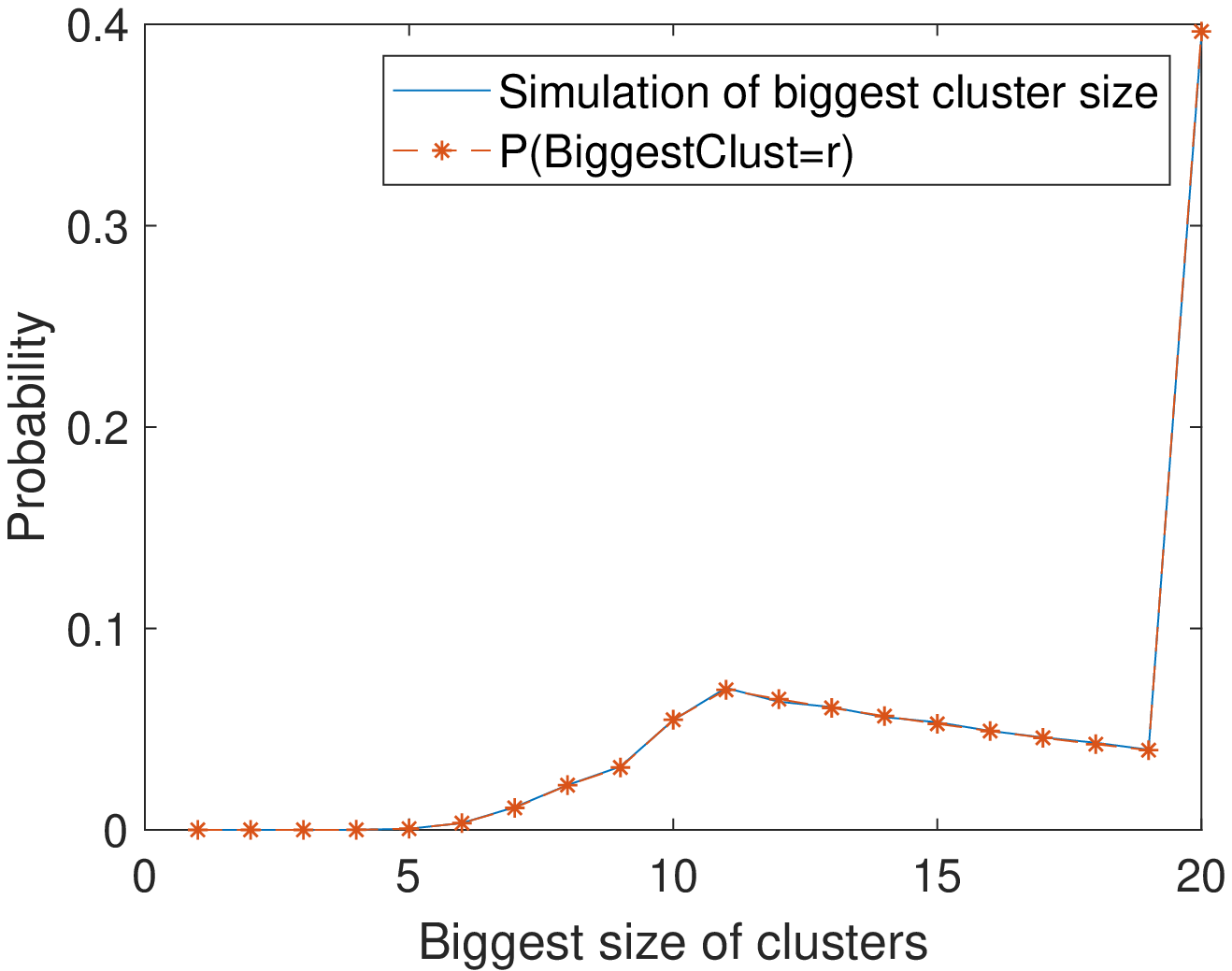}
  \caption{$n=20$,$\rho=0.05$}
  \label{fig:5}
\end{subfigure}
\caption{Probability distribution of size of the biggest cluster for different $n$ and $\rho$ }
\label{fig:images}
\end{figure}

Finally, the graphs of the distribution of the number of \textit{idle} vehicles are depicted on Figure 4. The simulation results are compared with the probabilities calculated by the formula (\ref{ttuu}). One can explain the graphs of distribution of \textit{idle} cars by the fact that in the case of a more stable connection (graphs \textbf{b} and \textbf{d}), the probability that \textit{idle} cars are not present in the network is close to 1, and then quickly decreases with increasing their number.
On graphs \textbf{a} and \textbf{c}, on the contrary, the probability of the network having several \textit{idle} vehicles is quite high due to the less stable connection. This can also be explained in terms of average numbers of \textit{idle} vehicle in cases \textbf{a}, \textbf{b}, \textbf{c}, and \textbf{d} equaling 2.4501, 0.1131, 4.4069, and 0.1357 respectively.



\begin{figure}
    \centering 
\begin{subfigure}{0.48\textwidth}\centering
  \includegraphics[width=\textwidth]{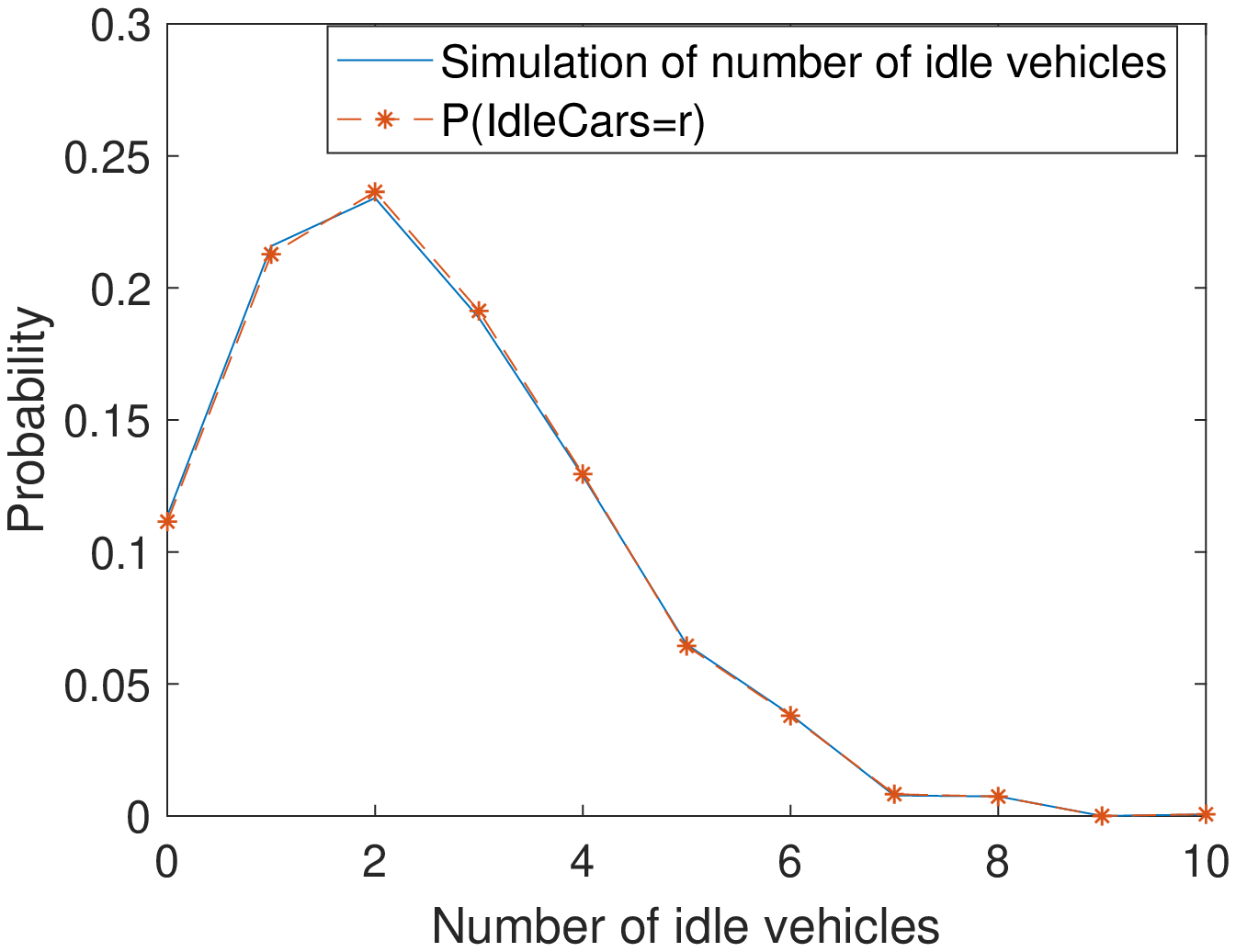}
  \caption{$n=10$,$\rho=0.01$}
  \label{fig:1}
\end{subfigure}\hfil 
\begin{subfigure}{0.48\textwidth}\centering
  \includegraphics[width=\textwidth]{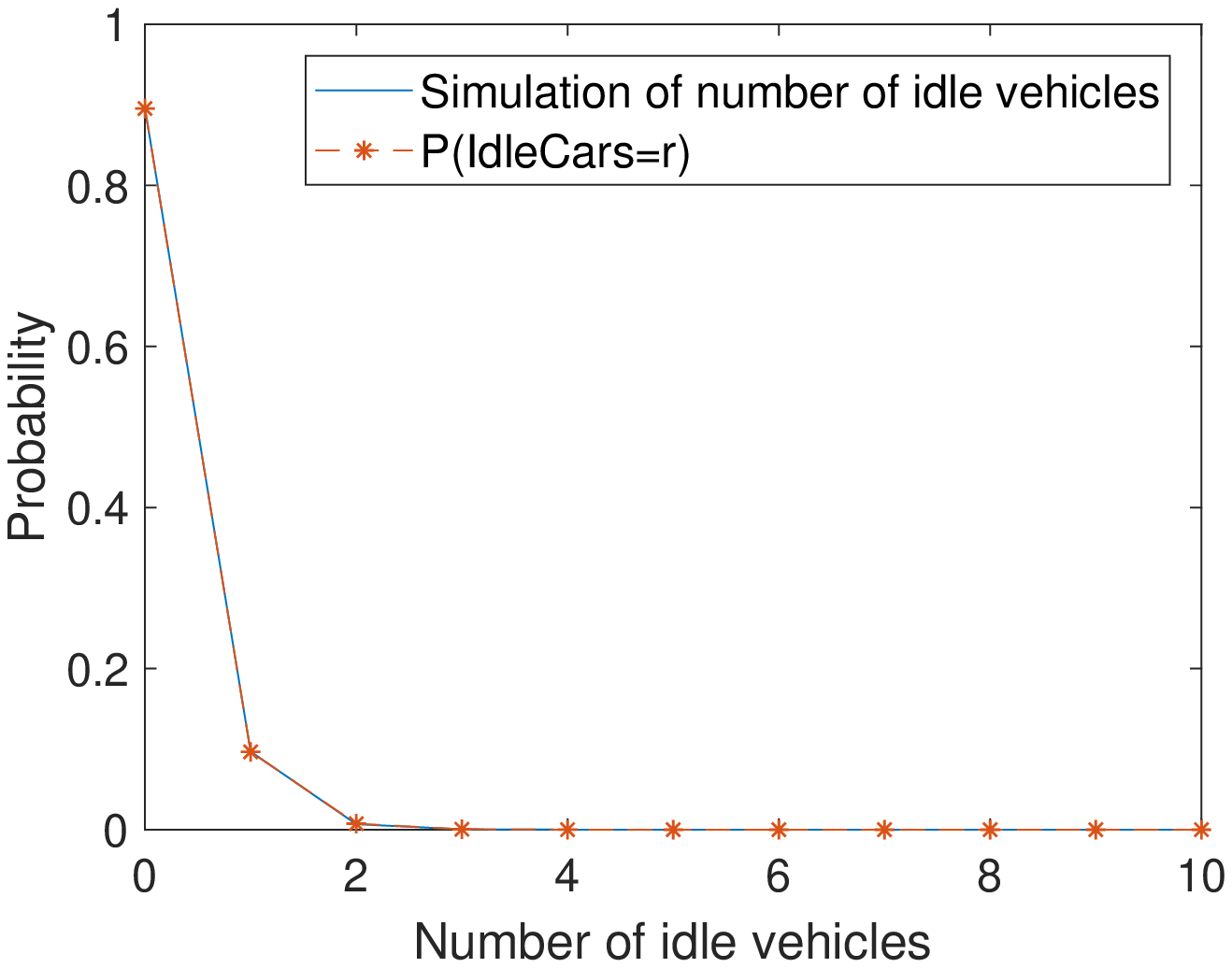}
  \caption{$n=10$,$\rho=0.05$}
  \label{fig:2}
\end{subfigure}

\medskip
\begin{subfigure}{0.48\textwidth}\centering
  \includegraphics[width=\textwidth]{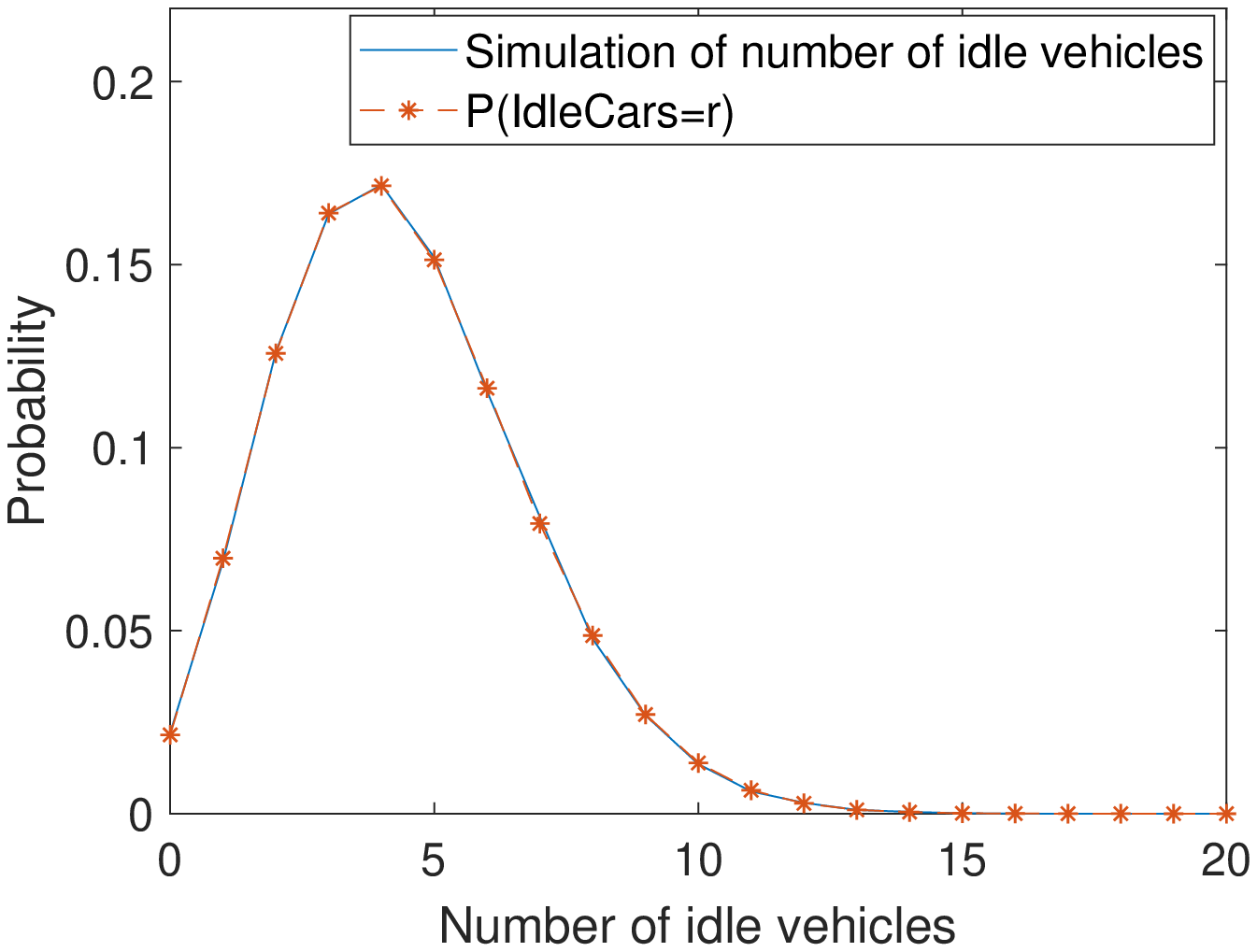}
  \caption{$n=20$,$\rho=0.01$}
  \label{fig:4}
\end{subfigure}\hfil 
\begin{subfigure}{0.48\textwidth}\centering
  \includegraphics[width=\textwidth]{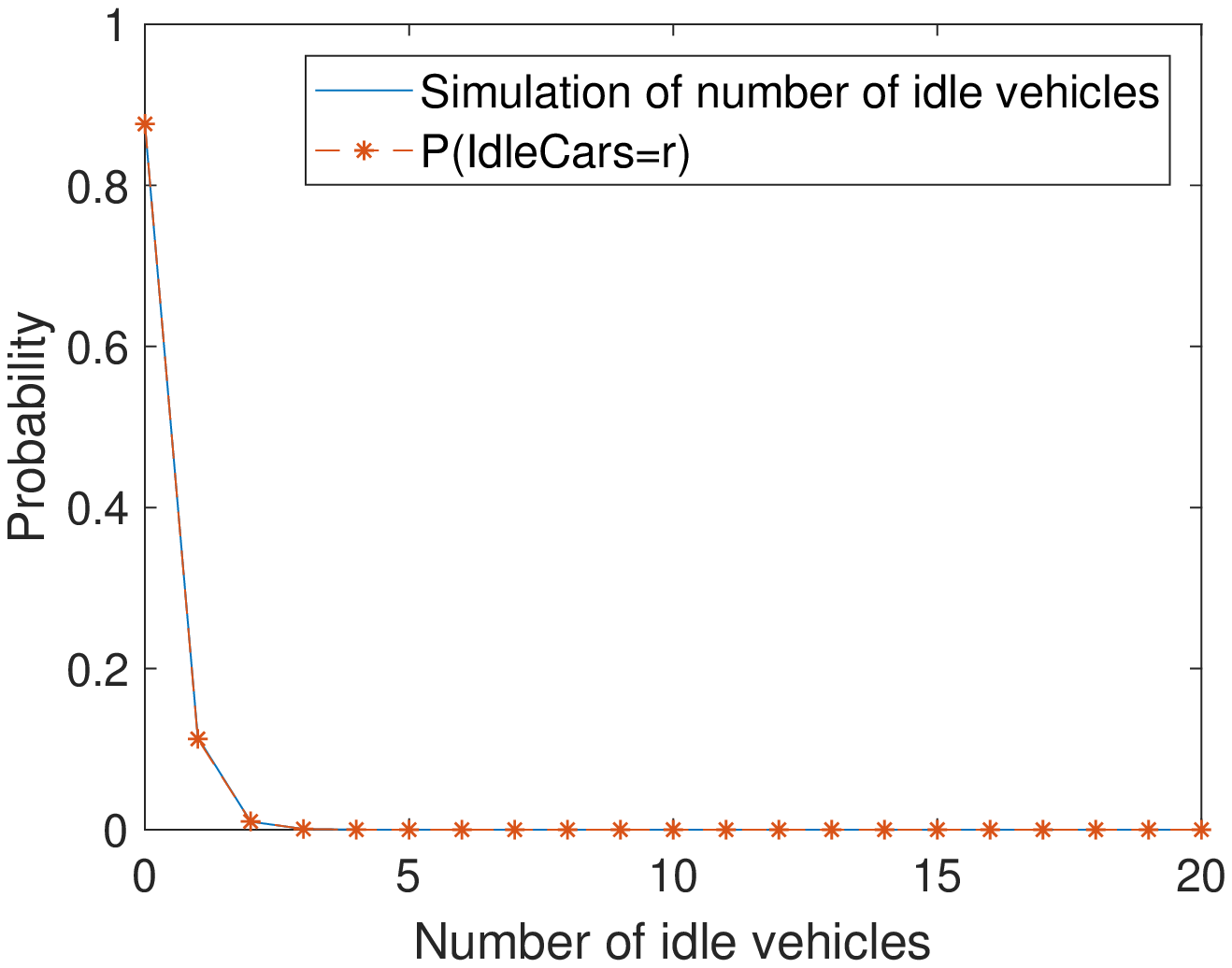}
  \caption{$n=20$,$\rho=0.05$}
  \label{fig:5}
\end{subfigure}
\caption{Probability distribution of the number of \textit{idle} vehicles for different $n$ and $\rho$ }
\label{fig:images}
\end{figure}

%

\section{Conclusion}
This article aims to present versatile investigation of VANET connectivity properties in the case, in which vehicles are distributed on the road. It expresses results in terms of parameters of known probability distributions of intervehicle distance and connectivity model. We derive distributions of number of clusters, cluster size, size of the biggest cluster and number of  \textit{idle} vehicles as well as calculate expected value and variance of every of these characteristics. The results are confirmed by the simulations in the cases of urban and rural traffic flow.


%
%
%
%
%
%
%

\appendix

\section{Method of generating functions} \label{Sec:2}

In this subsection we give a brief exposition of  mathematical method of generating functions. This elegant and effective method is used in this article to obtain distribution of cluster size. The essence of the method is that it
treats infinite sequence of numbers $a_k$ as the coefficients of a power series
$
\sum_{k=0}^\infty a_k x^k.
$
We use this method to establish a relation between the number of different representations of integer number as a sum of integer numbers with some restrictions. We touch only one aspect of this theory, for other applications we recommend to read the book \cite{Gener}. Here we formulate several problems in ascending order of complexity. We need the result of the problem 3, but in order to obtain it we solve problems 1 and 2.

\textbf{Problem 1.} \textit{What is the number of representations of positive integer number $n$ as a sum of $k$ positive integer summands?}

We denote coefficient of $x^n$ in power series $F(x)$ by $coeff_{x^n} F(x)$.
It is stated that the following coefficient:
\begin{equation}\label{121dwew}
  coeff_{x^n}\Big(x+x^2+x^3+\ldots\Big)^k
\end{equation}
equals the number of representations of $n$ as $k$ summands. Let us prove it. We can rewrite the sum $\Big(x+x^2+x^3+\ldots\Big)^k$ as
\begin{equation}
  \sum_{\alpha_1}\sum_{\alpha_2}\ldots \sum_{\alpha_k} x^{\alpha_1}x^{\alpha_2}\ldots x^{\alpha_k}=
  \sum_{\alpha_1}\sum_{\alpha_2}\ldots \sum_{\alpha_k} x^{\alpha_1+\alpha_2+\ldots+ \alpha_k}=
  \sum_{l=1}^\infty x^l \sum_{\alpha_1+\alpha_2+\ldots +\alpha_k =l}  1.
\end{equation}
Therefore, the coefficient (\ref{121dwew}) indeed equals the number of representations of $n$ as a sum of $k$ summands $\alpha_1,\alpha_2,\ldots,\alpha_k$.

\textbf{Problem 2.} \textit{ What is the number of representations of positive integer number $n$ as a sum of $k$ positive integer summands not equal $r$?}

The following formula gives the solution to this problem:
\begin{equation}\label{1212dwew}
  coeff_{x^n}\Big(x+x^2+x^3+\ldots -x^r\Big)^k.
\end{equation}
The only difference between formulas (\ref{121dwew}) and (\ref{1212dwew}) is that we subtract $x^r$ because integer summands do not equal $\alpha$. It can be proven similar to the proof of formula (\ref{121dwew}).

\textbf{Problem 3.} \textit{ What is the number of representations of positive integer number $n$ as a sum of $k$ positive integer summands with the following restriction: among them exactly $s$ numbers equal $r$?}

The answer is given by the following equality:
\begin{equation}\label{1212dwew1333}
  coeff_{x^n}\Big\{\binom{k}{s} x^{r s}\Big(x+x^2+x^3+\ldots -x^r\Big)^{k-s}\Big\}.
\end{equation}
because we can choose $s$ summands equaling $r$ in $\binom{k}{s}$ ways and other $k-s$ summands should not be equal $r$ (see problem~2).


\bibliography{article}%

\begin{thebibliography}{10}
\providecommand \doibase [0]{http://dx.doi.org/}%

\bibitem{R1}
{Hashem Eiza} M, {Owens} T, {Ni} Q. Secure and Robust Multi-Constrained QoS
  Aware Routing Algorithm for VANETs. {\it IEEE Trans. Dependable Secure
  Comput.} 2016\string; 13(1)\string: 32--45.

\bibitem{R2}
{Rak} J. LLA: A New Anypath Routing Scheme Providing Long Path Lifetime in
  VANETs. {\it IEEE Commun. Lett.} 2014\string; 18(2)\string: 281--284.

\bibitem{R3}
{Sahu} PK, {Wu} EH, {Sahoo} J, {Gerla} M. BAHG: Back-Bone-Assisted Hop Greedy
  Routing for VANET's City Environments. {\it IEEE Trans. Intell. Transp.
  Syst.} 2013\string; 14(1)\string: 199--213.

\bibitem{R4}
{Wu} Q, {Liu} Q, {Zhang} L, {Zhang} Z. A trusted routing protocol based on
  GeoDTN+Nav in VANET. {\it China Commun.} 2014\string; 11(14)\string:
  166--174.

\bibitem{R5}
{Eiza} MH, {Ni} Q. An Evolving Graph-Based Reliable Routing Scheme for VANETs.
  {\it IEEE Trans. Veh. Technol.} 2013\string; 62(4)\string: 1493--1504.

\bibitem{Cl1}
{Gong} H, {Yu} L, {Liu} N, {Zhang} X. Mobile content distribution with
  vehicular cloud in urban VANETs. {\it China Commun.} 2016\string;
  13(8)\string: 84--96.

\bibitem{Cl2}
{Salahuddin} MA, {Al-Fuqaha} A, {Guizani} M. Software-Defined Networking for
  RSU Clouds in Support of the Internet of Vehicles. {\it IEEE Internet Things
  J.} 2015\string; 2(2)\string: 133--144.

\bibitem{Cl3}
{Lin} C, {Deng} D, {Yao} C. Resource Allocation in Vehicular Cloud Computing
  Systems With Heterogeneous Vehicles and Roadside Units. {\it IEEE Internet
  Things J.} 2018\string; 5(5)\string: 3692--3700.

\bibitem{1}
Wang BX, Adams TM, Jin W, Meng Q. The process of information propagation in a
  traffic stream with a general vehicle headway: A revisit. {\it Transp.
  Research Emerg. Technol. {C}} 2010\string; 18(3)\string: 367 -- 375.
\newblock 11th IFAC Symposium: The Role of Control.

\bibitem{2}
Babu A, Muhammed~Ajeer V. Analytical model for connectivity of vehicular ad hoc
  networks in the presence of channel randomness. {\it Int. J. Commun. Syst.}
  2013\string; 26(7)\string: 927--946.

\bibitem{3}
{Kwon} S, {Kim} Y, {Shroff} NB. Analysis of Connectivity and Capacity in 1-D
  Vehicle-to-Vehicle Networks. {\it IEEE Trans. Wireless Commun.} 2016\string;
  15(12)\string: 8182--8194.

\bibitem{Clust}
{Wang} H, {Liu} RP, {Ni} W, {Chen} W, {Collings} IB. VANET Modeling and
  Clustering Design Under Practical Traffic, Channel and Mobility Conditions.
  {\it IEEE Trans. Wireless Commun.} 2015\string; 63(3)\string: 870--881.

\bibitem{ChModel}
Chandrasekharamenon NP, AnchareV B. Connectivity analysis of one-dimensional
  vehicular ad hoc networks in fading channels. {\it EURASIP Journal on
  Wireless Communications and Networking} 2012\string; 2012(1).

\bibitem{CarNet}
{Kesting} A, {Treiber} M, {Helbing} D. Connectivity Statistics of
  Store-and-Forward Intervehicle Communication. {\it IEEE Trans. Intell.
  Transp. Syst.} 2010\string; 11(1)\string: 172--181.

\bibitem{CarNet2}
{Yan} G, {Olariu} S. A Probabilistic Analysis of Link Duration in Vehicular Ad
  Hoc Networks. {\it IEEE Trans. Intell. Transp. Syst.} 2011\string;
  12(4)\string: 1227--1236.

\bibitem{Myev}
{Dubosarskii} G, {Primak} SL, {Wang} X. Evolution of Vehicle Network on a
  Highway. {\it IEEE Transactions on Vehicular Technology} 2019\string;
  68(9)\string: 9088--9097.

\bibitem{lr3}
Pekoz EA, Ross SM. A simple derivation of exact reliability formulas for linear
  and circular consecutive-k-of-n: F systems. {\it Journal of Applied
  Probability} 1995\string; 32(2)\string: 554–557.

\bibitem{LR}
Schilling MF. The Longest Run of Heads. {\it College Math. J.} 1990\string;
  21(3)\string: 196--207.

\bibitem{LR2}
Gordon L, Schilling MF, Waterman MS. An extreme value theory for long head
  runs. {\it Probab. Theory Relat. Fields} 1986\string; 72(2)\string: 279--287.

\bibitem{ChModel2}
{Rezha} FP, {Siadari} TS, Shin SY. Adaptive transmission power in cluster-based
  routing VANET. In: Jeju Island, Korea. ; 2012\string: 539--543.

\bibitem{Gener}
Wilf HS. {\it Generatingfunctionology}.
\newblock CRC Press, 3rd edition .
\newblock 2005.

\end{thebibliography}

\clearpage

%

\end{document}